\documentclass[12pt]{article}
\usepackage{fullpage}
\usepackage{graphicx,graphics}
\usepackage{amsmath,amsfonts,amssymb,amsthm}
\usepackage{hyperref}
\usepackage{color}
\usepackage[sort]{natbib}
\usepackage{subfigure}

\newcommand{\diag}{\mathrm{diag}}

\newcommand{\Tr}{\mathrm{tr}}

\newtheorem{theorem}{Theorem}

\newtheorem{proposition}{Proposition}

\newtheorem{corol}{Corollary}
\newtheorem{lemma}{Lemma}

\title{Regularized LRT for Large Scale Covariance Matrices: \\
One Sample Problem}
\author{Young-Geun Choi, Chi Tim Ng and Johan Lim\thanks{Young-Geun Choi and Johan Lim are at Department of Statistics, Seoul National University, Seoul, Korea. Chi Tim Ng is at Department of Statistics, 
Chonnam National University, Gwangju, Korea. All correspondences are to \texttt{johanlim@snu.ac.kr}}}
\date{\today}

\begin{document}
\maketitle

\begin{abstract}
\noindent The main theme of this paper is a modification of the likelihood ratio test (LRT)
for testing high dimensional covariance matrix.  {Recently, the 
correct asymptotic distribution of the LRT for a large-dimensional case (the case $p/n$ approaches to a constant $
\gamma \in (0,1]$) is specified by researchers. The correct procedure is named as 
corrected LRT. Despite of
its correction, the corrected LRT} is a function of sample eigenvalues 
that are suffered from redundant variability from high dimensionality and, subsequently, 
still does not have full power in differentiating hypotheses on the covariance matrix.
In this paper, motivated by the successes of a {linearly shrunken covariance matrix estimator} 
(simply shrinkage estimator) in various applications, we propose a regularized 
LRT that uses, in defining the LRT, the shrinkage estimator instead of the sample covariance matrix. 
We compute the asymptotic distribution of the regularized LRT, when the true covariance matrix 
is the identity matrix and a spiked covariance matrix. {The obtained asymptotic results have
applications in testing various hypotheses on the covariance matrix.} Here, we 
apply them to 
testing the identity of the true covariance matrix, 
which is a long standing problem in the literature, 
and show that the regularized LRT outperforms the corrected LRT, which is its 
non-regularized counterpart. In addition, we compare the power of the regularized LRT to
those of recent non-likelihood based procedures.
\end{abstract}

\baselineskip 20pt

\section{Introduction}\label{Intro}

{High dimensional data are now prevalent everywhere}
that include genomic data in biology, financial times
series data in economics, and natural language processing data in
machine learning and marketing. {The traditional 
procedures that assume that sample size $n$ is large and dimension
$p$ is fixed are not valid anymore for the analysis of high dimensional data.}  A significant amount of
research are made to resolve the difficulty from 
the dimensionality of the data.

This paper considers the inference problem of large scale covariance
matrix whose dimension $p$ is large compared to the sample size
$n$. To be specific, we are interested in testing whether the covariance matrix 
equals to a given matrix; $\mathcal{H}_0: \Sigma=\Sigma_0$, where $\Sigma_0$ can be set ${\rm I}_p$ without loss of generality. 
The likelihood ratio
test (LRT) statistic for testing $\mathcal{H}_0:\Sigma={\rm I}_p$ is defined by 
{
\begin{eqnarray} \nonumber
{\rm LRT}&=& \Tr \big( {\bf S}_n) - \log |{\bf S}_n| -p = \sum_{i=1}^p \big( l_i - \log l_i - 1  \big),
   \nonumber
\end{eqnarray}
where ${\bf S}_n$ is the unbiased and centered sample covariance matrix} and $l_i$ is the
$i-$th largest eigenvalue of the sample covariance matrix. 
{When $p$ is finite, LRT follows the chi-square distribution with degrees of freedom $p(p+1)/2$ asymptotically. However, this does not hold when $p$ increases.} Its correct asymptotic distribution is computed by \citet{Bai2009} for the 
case $p/n$ approaches $\gamma \in (0,1)$ and both $n$ and $p$ increase. They further numerically show that their asymptotic normal distribution defines a valid 
procedure for testing $\mathcal{H}_0: \Sigma={\rm I}_p$. 
The results of \citet{Bai2009} are refined by \citet{Jiang2012}, which include the asymptotic null distribution for the case $\gamma=1$. {Despite of the correction} of the null distribution, the sample 
covariance is known to have redundant variability when $p$ is large, and 
it still remains a general question that the LRT is asymptotically optimal for testing problem
in the $n,p$ large scheme.

{In this paper, it is shown that the corrected LRT can be further improved by introducing 
a linear shrinkage component.}
In detail, we consider a modification of the LRT, denoted by regularized LRT (rLRT), defined by
\begin{equation} \label{eqn:rlrt-def1}
{\rm rLRT}= \Tr \big( \widehat{\Sigma} \big) - \log |\widehat{\Sigma}| -p = 
 \sum_{i=1}^p \big( \psi_i - \log \psi_i - 1  \big),
\end{equation} 
where $\widehat{\Sigma}$ is a regularized covariance matrix and 
$\psi_i$ is the $i-$th largest eigenvalue of $\widehat{\Sigma}$. 
Here, we consider the regularization via linear shrinkage:
{
\begin{equation} \label{eqn:cov-lin}
\widehat{\Sigma} \equiv \lambda {\bf S}_n + (1- \lambda) {\rm I}_p.
\end{equation}
}
We also occasionally notate ${\rm rLRT}(\lambda)$ to emphasize the use of the value $\lambda$.
The linearly shrunken sample covariance matrix (simply shrinkage estimator) is
known to reduce expected estimation loss of the sample covariance matrix \citep{Ledoit2004}.
It is also successfully applied to many high-dimensional procedures to resolve the 
dimensionality problem. For example, \citet{Schafer2005} reconstruct a 
 gene regulatory network from microarray gene expression data using the 
inverse of a regularized covariance matrix. \citet{Chen2011} propose
a modified Hotelling's $T^2$-statistic for testing high dimensional mean vectors and apply 
it to finding differentially expressed gene sets. 
We are motivated by the success of above examples and inspect whether the power can be 
improved by the reduced variability via linear shrinkage. To the best of our knowledge, our work is 
the first time to apply the linear shrinkage to the covariance matrix testing problem itself.

We derive the asymptotic distribution of the proposed ${\rm rLRT}(\lambda)$ under two scenarios, 
(i) when $\Sigma = {\rm I}_p$ for the null distribution, and additionally (ii) when $\Sigma =
 \Sigma_{\rm spike}$ for power study. Here $\Sigma_{\rm spike}$ means a covariance matrix from the
 spiked population model \citep{Johnstone2001}, roughly it is defined as a covariance matrix whose
 eigenvalues are all $1$'s but some finite nonunit `spike'. The spiked covariance matrix 
assumed here includes the well known compound
 symmetry matrix $\Sigma_{\rm cs}(\rho)= {\rm I}_p + \rho {\rm J}_p$, 
where ${\rm J}_p$ is the $p \times p$ matrix of ones.
The main results show that  ${\rm rLRT}(\lambda)$ has normal distribution in asymptotic under both (i)
 and (ii); their asymptotic means are different but the variances are same. The main results are useful in
 testing various one sample covariance matrices. To be specific, first, in testing $\mathcal{H}_0:
 \Sigma={\rm I}_p$, (i) provides the asymptotic null distribution of ${\rm rLRT}(\lambda)$. Second,
 combining (i) and (ii)
 provides the asymptotic power for an arbitrary spiked alternative covariance matrix
 including $\Sigma_{\rm cs}(\rho)$. Finally, the results with $\lambda=1$ provide various asymptotic
 distributions of the corrected LRT. Among these many applications, in this paper, we particularly
focus on the LRT for testing $\mathcal{H}_0: \Sigma={\rm I}_p$, which has long been studied by many
 researchers \citep{Anderson2003,Ledoit2002,Bai2009,Chen2010,Jiang2012}.

The paper is organized as follows. In Section \ref{RMT}, we briefly review results of the random matrix theory
 that are essential to the asymptotic theory of the proposed rLRT. The results include the limit of empirical spectral distribution
 (ESD) of the sample covariance matrix and the central limit theorem (CLT) for linear spectral statistics
 (LSS). In Section \ref{MainRes}, we formally define the rLRT, and prove the asymptotic normality of the rLRT when 
the true covariance matrix $\Sigma$ is ${\rm I}_p$ or $\Sigma_{\rm spike}$. 
In Section \ref{Num}, the results developed in Section \ref{MainRes} are applied to testing  $\mathcal{H}_0: \Sigma={\rm I}_p$. Numerical study is provided to compare the powers of the LRT and other existing methods including the corrected LRT and other non-LRT tests by \cite{Ledoit2002} and \cite{Chen2010}.
In Section \ref{Disc}, we conclude the paper  with discussions of several technical details of the rLRT, for example,  close spiked eigenvalues.

\section{Random matrix theory}\label{RMT}

In this section, some useful properties of linear spectral statistics of the sample covariance matrix 
are introduced. The true covariance matrix $\Sigma$ is identity or that from a spiked population model.

The following notation is used throughout the paper. Let $\rm M$ be a  real-valued symmetric matrix of size $p \times p$ and $\alpha_j({\rm M})$ be the $j-$th largest
eigenvalue of the matrix $\rm M$ with natural labeling $\alpha_p({\rm M}) \le \cdots \le \alpha_1
({\rm M})$. The \emph{spectral distribution} (SD) for $\rm M$ is defined by
{
\begin{equation} \nonumber
F^{\rm M}(t) := \frac{1}{p}\sum_{j=1}^{p} \delta_{\alpha_j({\rm M})}(t), \quad t \in \mathbb{R},
\end{equation} 
}
where and $\delta_{\alpha}(t)$ is a point mass function that can be also written, with notational abuse, as $\delta_{\alpha}(t) = I(\alpha \le t)$. Here, $I(A)$ denotes the indicator function of a set $A$.

\subsection{Limiting spectral distribution of sample covariance matrix}

Let $\{z_{ij}\}_{i,j\geq 1}$ be an infinite double array of independent and identically distributed 
real-valued random variables with $Ez_{11}=0$, $Ez_{11}^2=1$ and $Ez_{11}^4=3$. Let ${\bf Z}_n = \{z_{ij}, i=1,2,\ldots,n, j=1,2,\ldots,p\}$ be
the top-left $n \times p$ block of the infinite double array. {We assume that
 both $n$ and $p$ diverge and their ratio $\gamma':= p/n$ converges to a positive constant $\gamma$.} The 
 data matrix and the {uncentered  sample covariance matrix} are ${\bf X}_n
= {\bf Z}_n\Sigma_p^{1/2}$ and ${\bf S}_n^0=\frac{1}{n}{\bf X}_n^{\top}{\bf X}_n$ respectively, 
where $\{ \Sigma_p, p=1,2,\ldots \}$ is a sequence of $p \times p$ nonrandom symmetric matrices.
Note that the fourth moment condition $Ez_{11}^4=3$ is used later on in Proposition \ref{Prop1}. 

In the random matrix theory literature \citep{Bai2004,Bai2009}, the limiting distribution of empirical SD 
{$F^{{\rm\bf S}_n^0}(\cdot)$} is determined by both the limits of $p/n$ and 
$F^{\Sigma_p}(\cdot)$. Specifically, if $H_p(\cdot) := F^{\Sigma_p}(\cdot)$ converges 
in distribution to $H(\cdot)$, a random distribution function $F^{{\rm\bf S}_n^0}(\cdot)$ converges in distribution
to a nonrandom distribution function, say $F^{\gamma, H}(\cdot)$ with probability one. The definition of $F^{\gamma, H}(\cdot)$ is given by its Stieltjes transform  
$m^{\gamma, H}(z)$ that is the unique solution of the following system of equations:
\begin{equation}\label{eqn:inv-stieltjes1}
m^{\gamma, H}(z)=\frac{1}{\gamma}\underline{m}^{\gamma, H}(z)+\frac{1-\gamma}{\gamma z} ; ~~
\end{equation}
\begin{equation}\label{eqn:inv-stieltjes}
z=-\frac{1}{\underline{m}^{\gamma, H}(z)}+\gamma \int \frac{t}{1+t\underline{m}^{\gamma, H}(z)}dH(t)
\end{equation}
on $z \in \{z : {\rm Im}(z)>0\}$.
Generally, $\underline{m}^{\gamma, H}(z)$ is known as the Stieltjes transform of 
the limiting SD of the so-called companion matrix for ${\bf S}_n^0$, which is defined by {$\underline{\bf S}_n^0:= \frac{1}{n}{\bf X}_n{\bf X}_n^{\top}$}. The density of $F^{\gamma, H}$ can be calculated from $\underline{m}^{\gamma, H}$ by the inversion formula,  
\begin{equation}\label{eqn:inv-stieltjes2}
\frac{dF^{\gamma, H}}{dx}(x)=\lim_{z\rightarrow x}~\frac{1}{\gamma\pi}~ {\rm Im}[\underline{m}^{\gamma, H}(z)], ~~x \in \mathbb{R}, ~~ z : {\rm Im}(z)>0.
\end{equation}

In the special case $\Sigma_p = {\rm I}_p$, we have $H_p(t) = \delta_1(t)$ and  
the corresponding spectral distribution $F^{\gamma,\delta_1}$ follows the Mar\u cenko-Pastur law.
To see this, note that the second equation of (\ref{eqn:inv-stieltjes}) can be rewritten as
\begin{equation}\label{eqn:inv-MPstieltjes}
z= - \frac{1}{\underline{m}^{\gamma, \delta_1}(z)} + \frac{\gamma}{1+\underline{m}^{\gamma, \delta_1}(z)},
\end{equation}
when $H=\delta_1$. {By the inversion formula, (\ref{eqn:inv-stieltjes2}) yields} the probability density function of the Mar\u cenko-Pastur distribution indexed by $\gamma$ when $0 < \gamma < 1$,
\begin{equation}\label{eqn:pdfMPlaw}
\frac{dF^{\gamma, \delta_1}}{dx}(x) = \frac{1}{2 \pi \gamma x} \sqrt{ \big( b(\gamma)-
x) (x- a (\gamma))}, \quad a (\gamma) \le x \le b(\gamma),
\end{equation}
where $a(\gamma):= (1-\sqrt{\gamma})^2 $ and $b(\gamma):= (1+\sqrt{\gamma})^2$.

\subsection{Central limit theorem for linear spectral statistics}

Many multivariate statistical procedures are based on {$F^{{\rm\bf S}_n}$}, 
the empirical SD of {the centered sample 
covariance matrix ${\rm\bf S}_n$.} Consider is a family of functionals of eigenvalues that is also
 called as linear spectral statistics (LSS) or linear eigenvalue statistics:
{
\[
 \frac{1}{p}\sum_{j=1}^{p}g(l_j) = \int g(x) d F^{{\bf S}_n}(x)
\]
}
where $g$ is a function fulfilling certain complex-analytic conditions.

{If the sample covariance matrix is uncentered, the central limit theorem (CLT) for the corresponding LSS  is developed in \cite{Bai2004} and \cite{Bai2009}. The proposition below is adapted from Theorem 2.1 in \cite{Bai2009} and used in the asymptotic property of the proposed regularized LRT under the null. 
Note that the centering term of the CLT 
possesses a \emph{finite-dimensional proxy} $F^{\gamma', H_p}(\cdot)$.}

\begin{proposition}[\cite{Bai2009}]\label{Prop1}
Let $T_n(g)$ be the functional
{
\begin{equation} \label{eqn:non-center}
T_n(g) = p \int g(x) \, d \! \left\{ \! F^{{\bf S}_n^0}(x) -F^{\gamma', H_p} (x) \! \right\}.
\end{equation}
}
Suppose that the two functions $g_{1}$ and $g_{2}\,,$ are complex analytic on an open domain containing an closed interval
$[ a(\gamma), b(\gamma) ]$ on the real axis. If $\gamma' = p/n \rightarrow \gamma \in (0, 1)$ and $H_p$ converges in distribution to $\delta_1$,
then the vector $(T_n(g_1),T_n(g_2))$ converges in distribution to a bivariate
normal distribution with mean 
\begin{eqnarray}
\mu(g_{i}) &=& \frac{g_{i}(a(\gamma))+g_{i}(b(\gamma))}{4}-\frac{1}{2
\pi}  \int_{a(\gamma)}^{b(\gamma)} \frac{g_{i}(x)}{\sqrt{4 \gamma
-(x-1- \gamma)^{2}}} dx \label{eqn:rmt_mean}
\end{eqnarray}
for $i=1,2,$ and variance
\begin{eqnarray}
v(g_{1},g_{2})=-\frac{1}{2\pi^{2}} \oint \oint
\frac{g_{1}(z_{1})g_{2}(z_{2})}{(\underline{m}(z_{1})-\underline{m}(z_{2}))^{2}}
d\underline{m}(z_{1}) d\underline{m}(z_{2}), \label{eqn:rmt_var}
\end{eqnarray}
where $\underline{m}=\underline{m}^{\gamma, \delta_1}$ is defined in 
(\ref{eqn:inv-MPstieltjes}). The two contours in (\ref{eqn:rmt_var}) are non-overlapping and
containing $[a(\gamma), b(\gamma)]$, the support of $F^{\gamma, \delta_1}$. 
\end{proposition}

{
The proposition requires the data to have known population mean vector (known as zero without loss
of generality) and considers 
the non-centered sample covariance matrix ${\bf S}_n^0$. However, this is seldomly true in 
practice and it is common to 
use the ``unbiased'' and ``centered'' sample covariance matrix ${\bf X}_n ={\rm B}_n {\bf Z}_n\Sigma_p^{1/2}$ 
and ${\bf S}_n = ({1}/\tilde{n}) {\bf X}_n^{\top} {\bf X}_n$, where $\tilde{n}=n-1$ and ${\rm B}_n = {\rm I}_n - ({1}/{n}) {\rm J}_n$. 
The extension of Proposition \ref{Prop1} to the centered sample covariance matrix is studied recently in \citet{Zheng2015}.
At the paper, the authors prove that, under the fourth moment assumptions (the assumption $Ez_{11}^4=3$ in Section 2.1), Proposition \ref{Prop1}
is still valid for ${\bf S}_n$ if $T_n(g)$ is redefined by
\begin{equation} \label{eqn:center}
T_n(g) = p \int g(x) \, d \! \left\{ \! F^{{\bf S}_n}(x) -F^{\tilde{\gamma}', H_p} (x) \! \right\},
\end{equation}
 where $\tilde{n}=n-1$ is the adjusted sample size and $\tilde{\gamma}'=p \big/ \tilde{n}$. This is named as ``substitution principle''.
}

{The assumption $H=\delta_1$ roughly implies that the spectrum of $ \Sigma_p $ is eventually concentrated around one. One simple example is $\Sigma_p = {\rm I}_p$. Then the SD is $H_p = F^{{\rm I}_p}=\delta_1$, which trivially converges to $\delta_1$.  In addition, we note that the spiked population model \citep{Johnstone2001} has $H=\delta_1$ as the limiting SD and is applicable to Proposition \ref{Prop1}. In our settings, the spiked population model refers to the data whose covariance matrix $\Sigma_p$ has the following eigenvalue structure:
\begin{eqnarray}
\underbrace{a_1, \ldots, a_1}_{n_1}, 
\underbrace{a_2, \ldots, a_2}_{n_2},  \ldots, \underbrace{a_k, \ldots, a_k}_{n_k}, \underbrace{1,  \ldots, 1}_{p - K}. & \label{eqn:spikedeigval}
\end{eqnarray}}
Then the SD $H_p$ corresponding to $\Sigma_p$ is
{
\begin{equation}\label{eqn:SDspike}
H_p (t) = \frac{p - K}{p}\delta_1(t) + \frac{1}{p}\sum_{i=1}^{K}{n_i  \delta_{a_i}(t)},
\end{equation}
}
where $K= n_1 + \ldots + n_k $ is a fixed finite integer, not depending on $n$ so that
 {$p - K$} eigenvalues of unity eventually dominate corresponding $H_p$ when $p$ is large. Thus, the limiting 
SD remains unchanged as $H=\delta_1$.

{To study the power of the proposed regularized LRT, it is useful to study how to apply the spiked population model to Proposition \ref{Prop1}. 
Although the limiting SD is simple ($\delta_1$), the spiked population model has 
several difficulties with the use of Proposition \ref{Prop1}.}  In the spiked model, its SD,
$H_p$ in (\ref{eqn:SDspike}), has masses at $K+1$ distinct points and 
$\underline{m}^{\tilde{\gamma}', H_p}(z)$ is the solution to a polynomial equation 
of degree $K+2$.  A polynomial equation has an analytical solution only when its degree
is less than or equal to 4. Therefore, if $K \geq 3$, we do not have an analytic
form of $\underline{m}^{\tilde{\gamma}', H_p}(z)$.
To resolve this difficulty, recently, \citet{Wang2013} provides an approximation formula of
$\int g dF^{\gamma', H_p}$ in Proposition \ref{Prop1} for the spiked population model. Such an approximation 
of $g(\underline{m}^{\gamma', H_p})$ at $\underline{m}^{\gamma', \delta_1}$ is constructed based on the idea  that $\underline{m}^{\gamma', H_p}(z)$ and  $\underline{m}^{\gamma', \delta_1}(z)$ would be close enough if $p$ is large. 

\begin{proposition}[\cite{Wang2013}]\label{Prop2}
Suppose that $\gamma' < 1$ and $H_p$ is given by (\ref{eqn:SDspike}),
with $|a_i - 1| > \sqrt{\gamma}$ for all $i=1,2,\ldots,K$. If a complex-valued function $g$ is analytic on an open domain containing the interval $[ a(\gamma), b(\gamma) ]$ and $k$ points $\varphi(a_i):= a_i + \frac{\gamma a_i}{a_i-1}$, $i=1,2,\ldots,K$ on the real axis, then $\int g(x) dF^{\gamma', H_p}(x)$ in (\ref{eqn:non-center}) can be approximated by
\begin{eqnarray}
&& \int g(x) dF^{\gamma', H_p}(x) \nonumber \\
&& \hspace*{-1cm} = \, - \frac{1}{2 \pi i p} \oint_{\mathcal{C}} g(-\frac{1}{\underline{m}}+\frac{\gamma'}{1+\underline{m}})
\left( \frac{K}{\gamma' \underline{m}} - \sum_{i=1}^{K} \frac{n_i a_i^2 \underline{m}}{(1 + a_i \underline{m})^2} \right) d\underline{m} \label{eqn:spiketerm1} \\
&& \hspace*{-1cm} + \frac{1}{2 \pi i p} \oint_{\mathcal{C}} f'(-\frac{1}{\underline{m}}+\frac{\gamma'}{1+\underline{m}}) 
\sum_{i=1}^{K}  \frac{(1-a_i)n_i}{(1+a_i \underline{m})(1+\underline{m})} \! \left( \frac{1}{\underline{m}} -\frac{\gamma' \underline{m}}{(1+\underline{m})^2} \! \right) d\underline{m} \label{eqn:spiketerm2} \\
&& \hspace*{-1cm} +\left(1 - \frac{K}{p}\right) \int g(x) d F^{\gamma', \delta_1}(x)+
\frac{1}{p}\sum_{i=1}^{K} n_i g(\varphi(a_i)) + O( \frac{1}{n^2} )\label{eqn:spiketerm3}
\end{eqnarray}
where $\underline{m}=\underline{m}^{\gamma', \delta_1}$ is defined in (\ref{eqn:inv-MPstieltjes}) by substituting $\gamma$ by $\gamma'$, and $\mathcal{C}$ is a counterclockwise contour enclosing the interval $\left[\frac{-1}{1-\sqrt{\gamma'}}, \frac{-1}{1+\sqrt{\gamma'}} \right]$ on the real axis.
\end{proposition}

The above proposition is a special case of Theorem 2 in \citet{Wang2013} when all the $a_i$'s are \emph{distant spikes}, i.e., $|a_i -1| > \sqrt{\gamma}$. Note that Theorem 2 of \cite{Wang2013} allows  \emph{close spikes} $a_i$ that is defined by $|a_i - 1| \leq \sqrt{\gamma}$. 
{In this paper, we focus on the alternative hypothesis with distant spike in the power study.}
{We finally remark that the substitution principle is directly applicable to Wang et al.'s results, that is, one can approximate $\int g(x) dF^{\tilde{\gamma}', H_p}(x)$ in (\ref{eqn:center}) by Proposition \ref{Prop2}, with $\gamma' = p/n$ in the formula replaced by $\tilde{\gamma}' = p/(n-1)$.}

\section{Main results}\label{MainRes} 

In this section, the asymptotic results of the rLRT are presented. Here,
 the rLRT is defined via the linear shrinkage estimator instead of the sample covariance matrix :
\begin{equation} \nonumber
{\rm rLRT} (\lambda) := \Tr \big( \widehat{\Sigma} \big)  -
	\log \big| \widehat{\Sigma} \big| -p, 
	~~\mbox{where}~  \widehat{\Sigma} := \lambda {\bf S}_n + (1- \lambda) {\rm I}_p.
\end{equation}
{The shrinkage intensity  $\lambda$ is fixed and chosen from $(0,1)$. Define 
$\psi(x)= \lambda x + (1-\lambda)$ and $g(x) = \psi(x) - \log\{ \psi(x)\} -1$.
We consider ${\rm rLRT}(\lambda) = p \int g(x) dF^{{\bf S}_n}(x)$, whose sample covariance matrix term is of the centered version. Then Proposition \ref{Prop1} along with the substitution principle
yields the following results.} 

\begin{theorem} \label{Thm1} 
Let $g(x) = \psi(x) - \log\{ \psi(x)\} -1$ and $\psi(x)= \lambda x + (1-\lambda)$ with fixed $\lambda \in (0,1)$. Suppose that $\Sigma_p={\rm I}_{p}$. If {$\tilde{\gamma}'= p\big/\tilde{n}\rightarrow \gamma  \in (0,1)$ with $\tilde{n}=n-1$, then
\begin{equation} \nonumber
T_n(g) = {\rm rLRT}(\lambda) -p \int g(x)d F^{\tilde{\gamma}', \delta_1} (x)
\end{equation}
}converges in distribution to the normal distribution with mean
\begin{equation}
\mu(g) =  -\frac{\log\sqrt{(1+\lambda\gamma)^2-4\lambda^2\gamma}}{2}+\frac{1}{4\pi}\int^{2\pi}_{0} \left( \log(1+\lambda\gamma-2\lambda\sqrt{\gamma}\cos\theta) \right)d\theta \label{eqn:a_mean}
\end{equation}
and variance
\begin{equation}
v(g) = 2\left\{ -\frac{\lambda}{M}-\lambda(1+\gamma-\lambda\gamma)+\frac{\lambda\gamma}{1+N}-\log\frac{M-N}{M(1+N)} \right\}, \label{eqn:a_var}
\end{equation}
where
\begin{equation}\label{MN}
 M,N= M(\lambda, \gamma), N(\lambda, \gamma):= \frac{-(1-2\lambda+\lambda\gamma)\pm\sqrt{(1-2\lambda+\lambda\gamma)^2+4\lambda(1-\lambda)}}{2(1-\lambda)}
\end{equation}
\end{theorem}
\noindent The detailed proof of Theorem \ref{Thm1} is given in Appendix \ref{Pf}.

{Note that when $\lambda=1$, $\mu(g)=-\log(1-\gamma)/2$ and $v(g)=-2\gamma-2\log(1-\gamma)$ that are consistent with the results in \cite{Bai2009}. To see this, observe that the integral in the mean function
$
\int^{2\pi}_{0}  \log \big(1+\lambda\gamma-2\lambda\sqrt{\gamma}\cos\theta \big) d\theta
$
approaches zero according to the dominate convergence theorem.} In addition, it can be shown that $M$ goes to $-1/(1-\gamma)$, and $N$ goes to $+\infty$ as $\lambda\rightarrow 1$ using 
the approximation formula $\sqrt{x+\Delta x}\approx \sqrt{x}+\frac{1}{2}x^{-1/2} \Delta x$.

{
Next, consider the finite-dimensional proxy 
\begin{equation} \nonumber 
\int g(x)dF^{\tilde{\gamma}', \delta_1}(x).
\end{equation}
}
From the density function of Mar\u{c}enko-Pastur law (\ref{eqn:pdfMPlaw}) and the fact that $1 = \int x dF^{\tilde{\gamma}', \delta_1}(x) = \int 1 dF^{\tilde{\gamma}', \delta_1}(x)$, 
\begin{eqnarray*}
&& \int g(x)dF^{\tilde{\gamma}', \delta_1}(x) \\
&=&  \int_{a(\tilde{\gamma}')}^{b(\tilde{\gamma}')} \frac{(\lambda x-\lambda)-\log(\lambda x+1-\lambda)}{2\pi x\tilde{\gamma}'} \sqrt{\{b(\tilde{\gamma}')-x \} \{x-a(\tilde{\gamma}') \}}dx \\
&=& - \int_{a(\tilde{\gamma}')}^{b(\tilde{\gamma}')} \frac{\log(\lambda x+1-\lambda)}{2\pi x\tilde{\gamma}'} \sqrt{\{b(\tilde{\gamma}')-x \} \{x-a(\tilde{\gamma}') \}}dx.
\end{eqnarray*}
By substituting  $x=1+\tilde{\gamma}'-2\sqrt{\tilde{\gamma}'}\cos\theta$,
we have an alternative representation of the integral
\begin{equation}\label{Thm1Impl}
\int g(x)dF^{\tilde{\gamma}', \delta_1}(x)=  -
\frac{2}{\pi }\int^{\pi}_0 \frac{\log(1+\lambda\tilde{\gamma}'-2\lambda\sqrt{\tilde{\gamma}'}\cos\theta)}{1+\tilde{\gamma}'-2\sqrt{\tilde{\gamma}'}\cos\theta} \cdot  \sin^2\theta d\theta.
\end{equation}
It is remarked that the right-hand-side of (\ref{Thm1Impl}) can be evaluated via the standard numerical integration techniques.

Theorem \ref{Thm2} below establishes the asymptotic normality of the rLRT under the alternative hypothesis that the true covariance matrix from the spiked population model in Section \ref{RMT}. It follows directly from Proposition \ref{Prop2}.
\begin{theorem} \label{Thm2} 
Let $g(x) = \psi(x) - \log\{ \psi(x)\} -1$ and $\psi(x)= \lambda x + (1-\lambda)$ with fixed $\lambda \in (0,1)$. Suppose that $\Sigma_p$ has SD {$H_p (t) = \frac{p - K}{p}\delta_1(t) + \frac{1}{p}\sum_{i=1}^{K}{n_i \delta_{a_i}(t)}$} as in (\ref{eqn:SDspike}) with $|a_i - 1| > \sqrt{\gamma}$ for all $i=1,2,\ldots,K$.
If {$\tilde{\gamma}'= p/\tilde{n} \rightarrow \gamma  \in (0,1)$ with $\tilde{n}=n-1$}, then
\begin{equation} \nonumber 
{\rm rLRT(\lambda)} - p \int g(x) dF^{\tilde{\gamma}', H_p}(x)  \longrightarrow N \big( \mu(g), v(g) \big),
\end{equation} 
where $\mu(g),v(g)$ are defined in Theorem \ref{Thm1} and
\begin{equation} \nonumber 
 {p \int g(x) dF^{\tilde{\gamma}', H_p}(x)}
= \left( 1 - \frac{K}{p} \right)  {\int g(x) dF^{\tilde{\gamma}', \delta_1}(x)} + \frac{K}{p}{C (\lambda, \tilde{\gamma}')}  + O\left(\frac{1}{n^2}\right)
\end{equation} 
with
\begin{eqnarray*}
{C (\lambda, \tilde{\gamma}')} &=& \lambda \cdot \frac{1}{K} \sum_{i=1}^{K} n_i a_i - \lambda  -  \frac{1}{K}\sum_{i=1}^{K} n_i \log \psi \{ \varphi(a_i) \} \\
&& \hspace*{-1.2cm} - \Bigg[ \frac{1}{ {\tilde{\gamma}'} } \log(-M) + \frac{1}{K}\sum_{i=1}^{K}n_i \log \Big( \frac{1-a_i}{1 + a_i M} \Big) 
- \frac{1}{K} \sum_{i=1}^{K} n_i \Big( \frac{1}{1 + a_i M} - \frac{1}{1 - a_i} \Big)  \Bigg] \\
&& \hspace*{-1.2cm} + \frac{\lambda}{(1 - \lambda)K} \sum_{i=1}^{K} n_i \Bigg[ \frac{1}{M-N} \Bigg\{ \frac{a_i (M+1)}{1+ a_i M} -\frac{a_i {\tilde{\gamma}'} M^2}{(1 + a_i M)(M+1)} - 1 + \frac{{\tilde{\gamma}'} M^2}{(M+1)^2} \Bigg\} 
\\
&& \hspace*{-1.2cm}~~~~~~~~~~~~~\qquad\qquad - \frac{1}{(M+1)(N+1)} \Bigg\{ \frac{a_i {\tilde{\gamma}'}}{1 - a_i} + \frac{{\tilde{\gamma}'} (2MN + M + N)}{(M+1)(N+1)} \Bigg\} \Bigg],
\end{eqnarray*}
where $\varphi(a)= a + \frac{\gamma a}{a-1}$ and both $M$ and $N$ in $C (\lambda, \tilde{\gamma}')$ are $M = M(\lambda, \tilde{\gamma}')$ and $N = N(\lambda, \tilde{\gamma}')$.
\end{theorem}
\noindent The proof of the above theorem is provided in Appendix \ref{Pf}.

{Theorem \ref{Thm2} can be applied to {$\Sigma_{{\rm cs}}$}, the covariance matrix with compound symmetry, which is defined by}
\begin{equation}\label{cpsymm}
\Sigma_{{\rm cs}} \left( \frac{\beta}{p} \right) = \left(\begin{array}{ccccc}1 + \beta/p &\beta/p&\beta/p&\cdots&\beta/p\\
\beta/p&1+\beta/p&&&\\
\vdots&&\ddots&&\vdots\\
\vdots&&&\ddots&\vdots\\
\beta/p&\cdots&\cdots&\cdots&1+ \beta/p
\end{array}\right) = {\rm I}_{p} + \frac{\beta}{p} {\rm J}_p.
\end{equation}
This matrix has a spiked eigenvalue structure; $1 + \beta$ for one eigenvalue and $1$ for the other $p-1$ eigenvalues. The corresponding SD is $H_p(t)=\frac{p -1}{p} \delta_1(t) + \frac{1}{p}\delta_{1+\beta}(t)$. Theorem \ref{Thm2} with $K=1$, $k=1$, $n_1=1$, and $a_1 = 1 + \beta$ gives the following corollary.

\begin{corol} \label{Cor1}
Let $g(x) = \psi(x) - \log\{ \psi(x)\} -1$ and $\psi(x)= \lambda x + (1-\lambda)$ with fixed $\lambda \in (0,1)$. Suppose that $\Sigma_p$ has SD $H_p(t)=\frac{p -1}{p} \delta_1(t) + \frac{1}{p}\delta_{1+\beta}(t)$ with $\beta > \sqrt{\gamma}$. If {$\tilde{\gamma}' := p/\tilde{n} \rightarrow \gamma  \in (0,1)$ where $\tilde{n}=n-1$}, then
\begin{equation} \nonumber 
{\rm rLRT(\lambda)} - {p \int g(x) dF^{\tilde{\gamma}', H_p}(x)}  \longrightarrow N( \mu(g), v(g)),
\end{equation} 
where $\mu(g),v(g)$ are defined in Theorem \ref{Thm1} and
\begin{equation}\nonumber
{ \int g(x) dF^{\tilde{\gamma}', H_p}(x) }
= \left(1 - \frac{1}{p}\right) { \int g(x) dF^{\tilde{\gamma}', \delta_1}(x) } + \frac{1}{p} {C (\lambda, \tilde{\gamma}') } + O\left(\frac{1}{n^2} \right)
\end{equation}
with
\begin{eqnarray*}
{C (\lambda, \tilde{\gamma}')} &=& \lambda\beta - \log \psi \{ \varphi(1+\beta) \} + \frac{1}{ { \tilde{\gamma}'} } \left\{ \frac{1}{\lambda} + \log(-\frac{1}{M}) \right\}  \\
&& \hspace*{-1.2cm} + \frac{1}{1 + (1+\beta) M} -   \log \Big( -\frac{\beta}{1 + (1+\beta)M} \Big) \\
&& \hspace*{-1.2cm} + \frac{\lambda}{(1 - \lambda)(M-N)} \! \Bigg\{ \! \frac{(1+\beta) (M+1)}{1+ (1+\beta) M} \! - \! \frac{ {\tilde{\gamma}'} (1+\beta) M^2}{(1 + (1+\beta) M)(M+1)} \! - \! 1\! +\! \frac{{\tilde{\gamma}'} M^2}{(M+1)^2}\! \Bigg\},
\end{eqnarray*}
where $\varphi(a)= a + \frac{\gamma a}{a-1}$ and both $M$ and $N$ in $C (\lambda, \tilde{\gamma}')$ are $M = M(\lambda, \tilde{\gamma}')$ and $N = N(\lambda, \tilde{\gamma}')$.
\end{corol}
\noindent In the corollary, the condition $\beta > \sqrt{\gamma}$ means that the spiked eigenvalue 
$1 + \beta$ is distant.

\section{Testing the identity covariance matrix}\label{Num} 

The results in Section \ref{MainRes} can be used for testing various hypotheses on one 
sample covariance matrix. In this section, we study the finite-sample properties of the proposed rLRT in testing $\mathcal{H}_0:
\Sigma={\rm I}_p$. Additionally, we compare the power of the proposed rLRT and the following existing procedures in the literature. 

\begin{itemize}
\item \cite{Ledoit2002} assume that $p\big/n \rightarrow \gamma \in (0, \infty)$ and propose a statistic
\begin{equation} \nonumber 
{\bf T}_{\rm LW} = \frac{1}{p} {\rm tr} \left\{ \big( {\bf S} - {\rm I} \big)^2 \right\} 
-\frac{p}{n} \left\{ \frac{1}{p} {\rm tr} \big( {\bf S} \big) \right\}^2 + \frac{p}{n}.
\end{equation} 
The asymptotic distribution of ${\bf T}_{\rm LW}$ is, if both 
$n$ and $p$ increase with $p/n \rightarrow \gamma \in (0, \infty)$, 
\begin{equation} \nonumber 
n {\bf T}_{\rm LW} - p
\end{equation} 
converges in distribution to normal distribution with mean $1$ and variance $4$.

\item {\cite{Bai2009} and \citet{Zheng2015} propose a corrected LRT for the cases where both $n$ and $p$ increases and $p\big/n$ converges to $\gamma \in (0,1)$. The corrected LRT statistic is 
\begin{equation} \nonumber 
{\rm cLRT} =   \Tr \big({\rm \bf S}_n\big) - \log \big| {\rm\bf S}_n \big| -p.
\end{equation}
They show that
\begin{equation} \nonumber 
{\bf T}_{\rm lrt} = v(g)^{-1/2} \bigg\{  {\rm cLRT} - p \int g(x) dF^{\tilde{\gamma}', \delta_1}(x)dx -\mu (g) \bigg\} 
\end{equation}
converges in distribution to the standard normal distribution, where $\mu(g) = - \frac{\log (1- \gamma)}{2}$, $v(g) = - 2 \log(1-\gamma) - 2\gamma$, and $\int g(x) dF^{\tilde{\gamma}', \delta_1}(x)dx = 1 - \frac{\tilde{\gamma}'-1}{\tilde{\gamma}'} \log \big(1-\tilde{\gamma}' \big)$.
}

\item Finally, \cite{Chen2010} proposes to use the statistic 
\begin{equation} \nonumber 
{\bf T}_{\rm C} = \frac{1}{p} {\bf V}_{2.n} - \frac{2}{p} {\bf V}_{1.n} +1,
\end{equation}
where 
\begin{eqnarray}
{\bf V}_{1.n}&=& \frac{1}{n} \sum_{i=1}^n X_i^{\top} X_i - \frac{1}{P_n^2} \sum_{i \neq j} X_i^{\top} X_j \nonumber\\
{\bf V}_{2.n} &=&  \frac{1}{ P_n^2} \sum_{i \neq j} \big( X_i^{\top} X_j \big)^2 - 
2 \frac{1}{P_n^3} \sum_{i,j,k}^* X_i^{\top} X_j X_j^{\top} X_k \nonumber \\
&& + \frac{1}{P_n^4} \sum_{i,j,k,l}^* X_i^{\top} X_j X_k^{\top} X_l \nonumber\\
P_n^r &=& n!/(n-r)! \nonumber
\end{eqnarray}
and $\sum^*$ is the summation over different indices.  
The asymptotic theory suggests that, under the null, $n {\bf T}_{\rm C}$
converges in distribution to the normal distribution with mean $0$ and variance $4$. 
\end{itemize}

\subsection{Power comparison with the cLRT}\label{Num1}

The asymptotic power curves of the cLRT and rLRT 
for the alternative hypothesis of the compound 
symmetry can be obtained using Corollary \ref{Cor1}. When $\Sigma_p \equiv \Sigma_{\rm cs} (\beta/p)$ and  $\tilde{\gamma}' \rightarrow \gamma$, the probability of rejecting $\mathcal{H}_0 : \Sigma_p= {\rm I}_p$ at level $\eta$ is
\begin{equation}\label{eqn:CSpower}
1 - \Phi\left[  \Phi^{-1}(\eta) +{\frac{1}{v(g)}\left( \int g(x) dF^{\tilde{\gamma}', \delta_1}(x) - C (\lambda, \tilde{\gamma}') \right)} \right], ~~\beta > \sqrt{\gamma}
\end{equation}
where ${C (\lambda, \tilde{\gamma}')}$ is defined in Corollary \ref{Cor1} and $\Phi(\cdot)$ denotes the cumulative distribution function of the standard normal distribution.

\begin{figure}[ht!]
\centering
\begin{minipage}{.45\textwidth}
  \centering
  \includegraphics[width=1.0\linewidth]{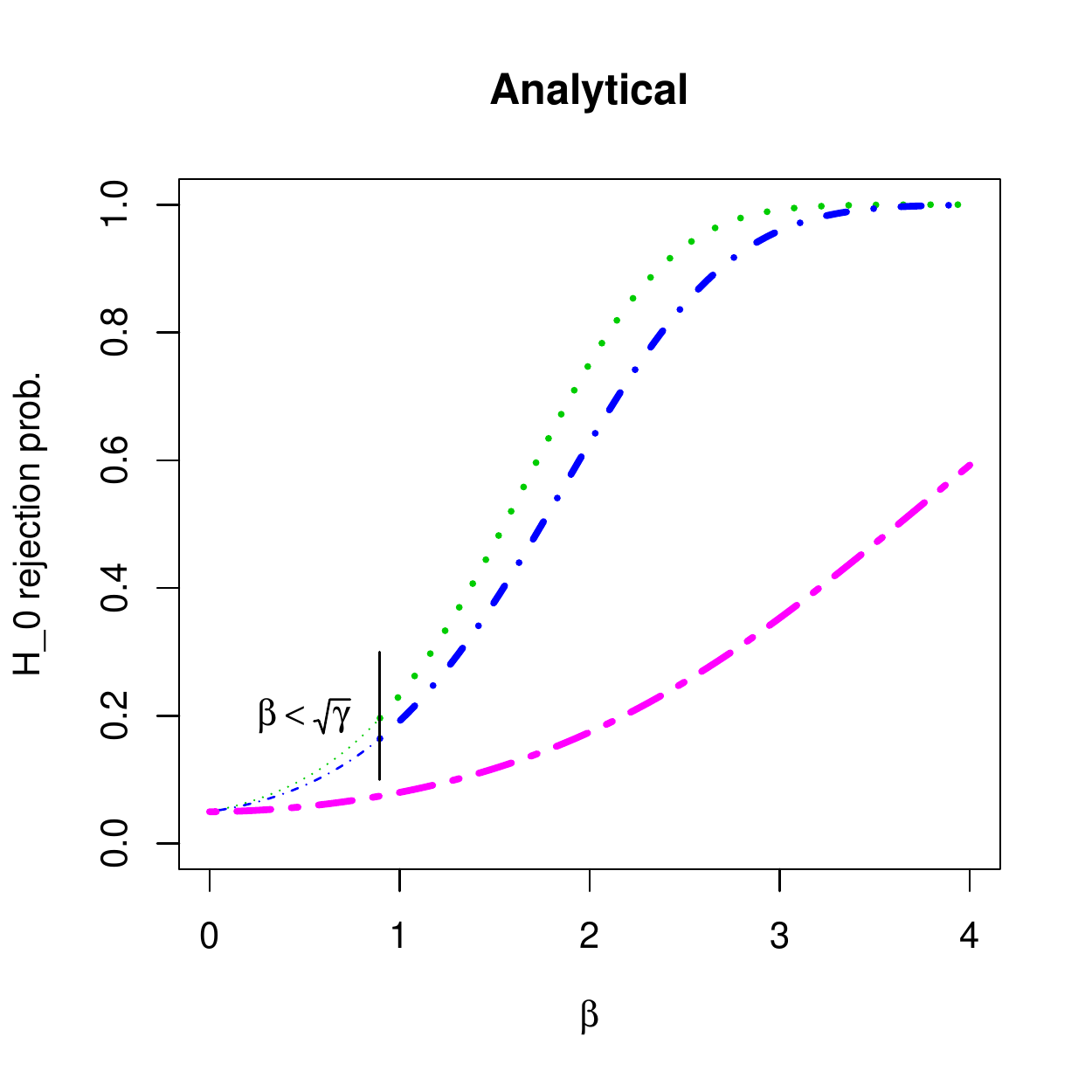}
\end{minipage}%
\begin{minipage}{.45\textwidth}
  \centering
  \includegraphics[width=1.0\linewidth]{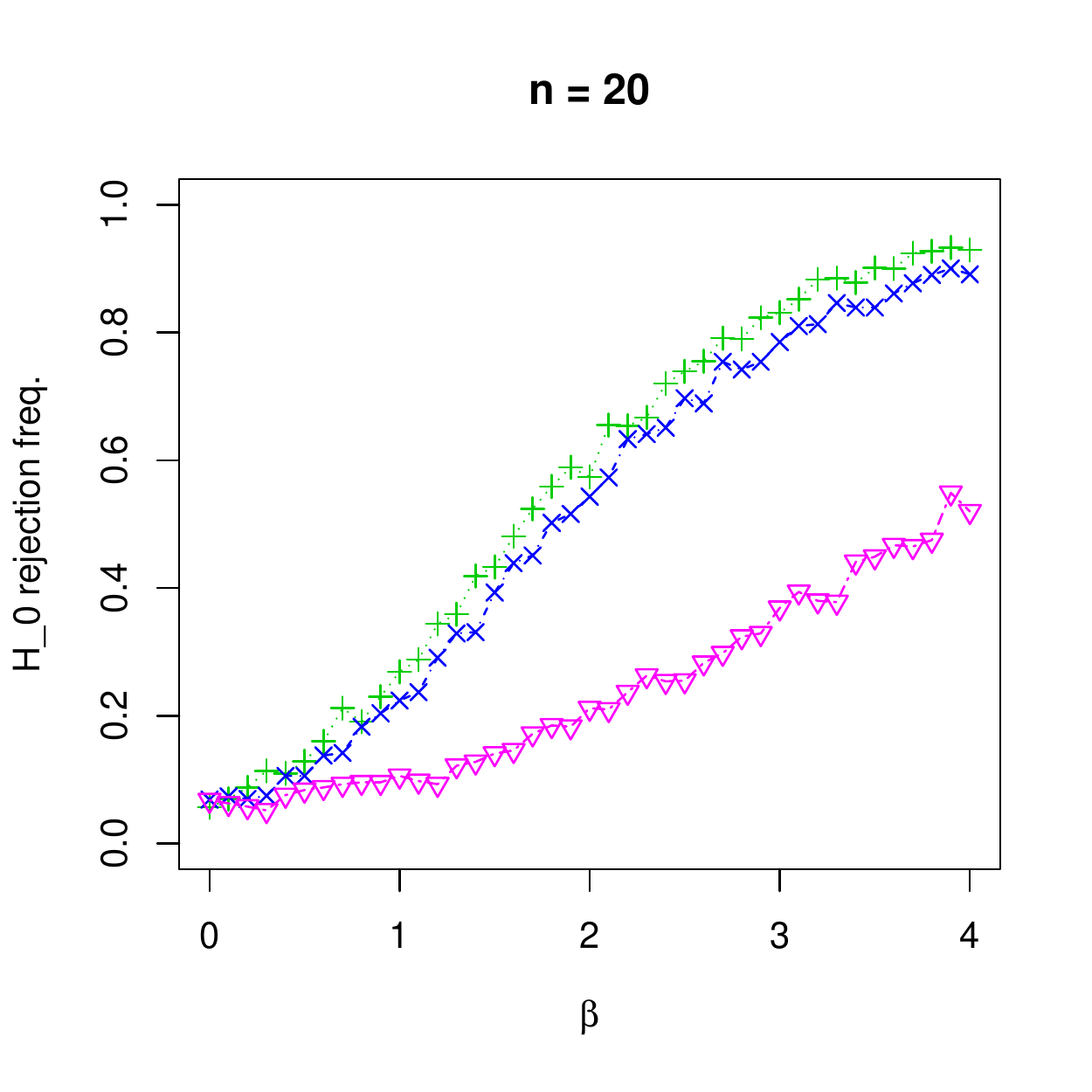}
\end{minipage}
\begin{minipage}{.45\textwidth}
  \centering
  \includegraphics[width=1.0\linewidth]{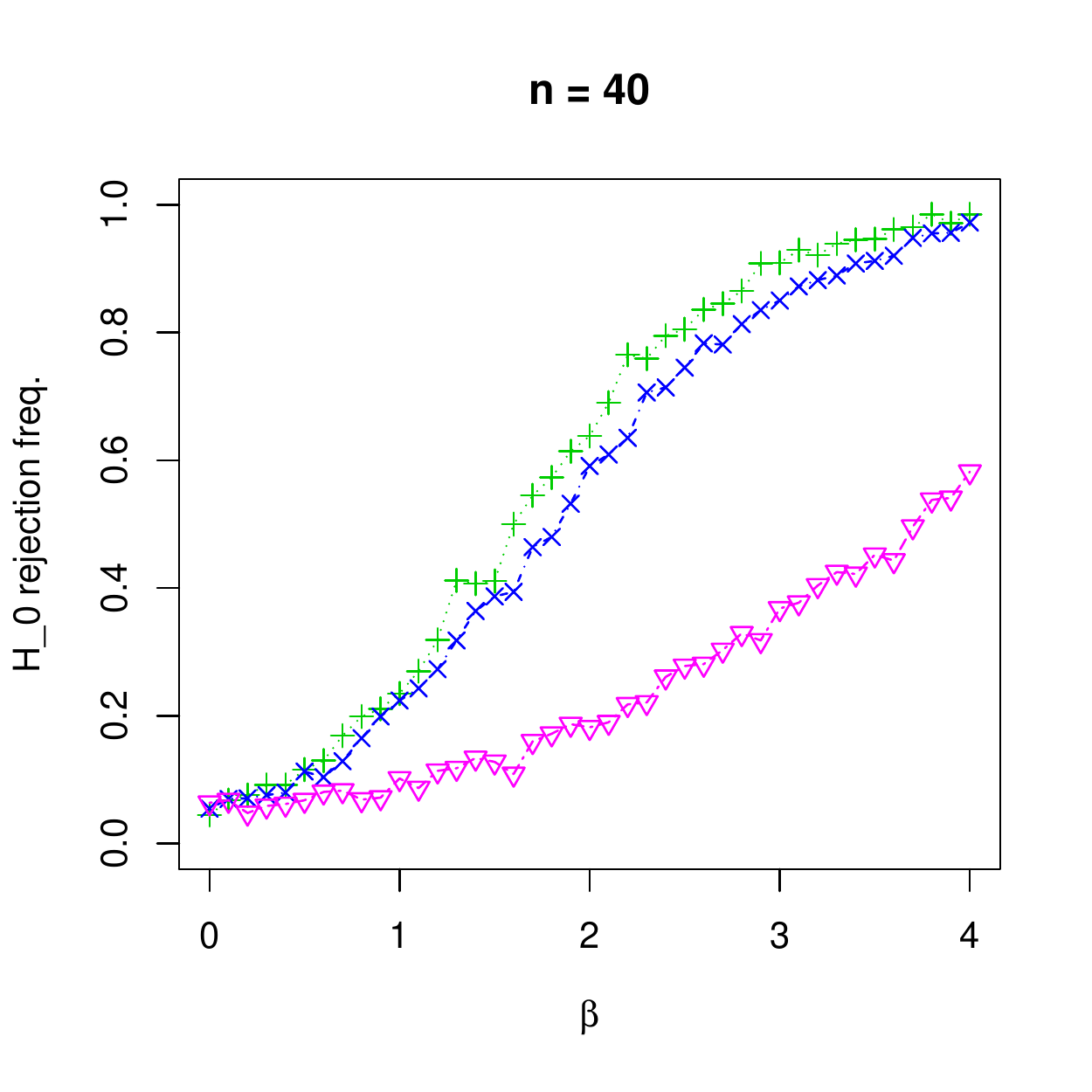}
\end{minipage}%
\begin{minipage}{.45\textwidth}
  \centering
  \includegraphics[width=1.0\linewidth]{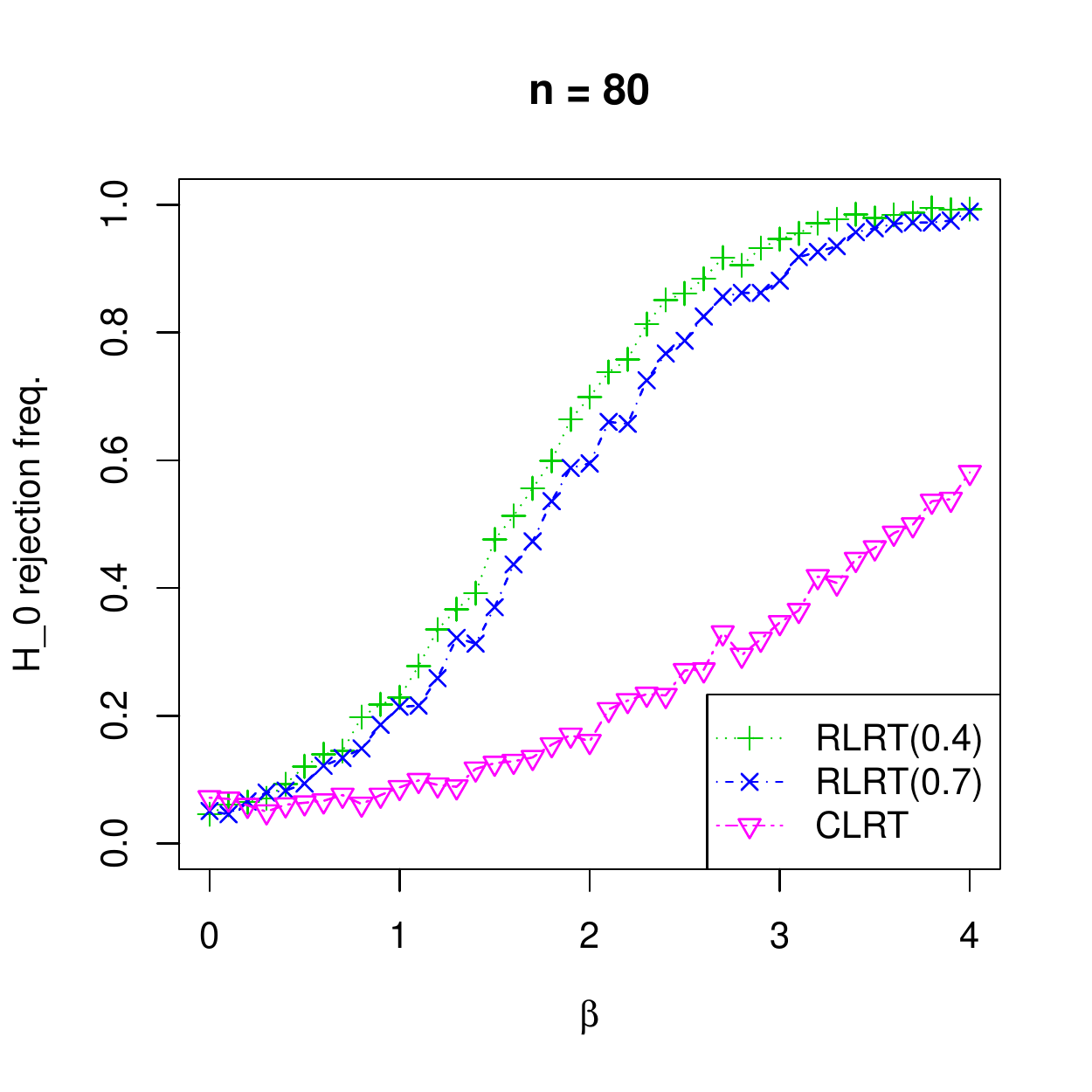}
\end{minipage}
\caption{Analytic and empirical power curves for the rLRT and cLRT.}
\label{fig:powerCS}
\end{figure}

The powers of the cLRT and rLRT with $\lambda=0.4$ and $\lambda=0.7$ are plotted in Figure \ref{fig:powerCS}. Each panel of Figure \ref{fig:powerCS} compares the powers of the cLRT and rLRT 
for different sample size $n$. In each panel, the results of $\beta<\sqrt{\gamma}$ (close spike) are also included to study the performances when the the assumption of distant spike in Corollary \ref{Cor1} is violated. The results
 in Figure \ref{fig:powerCS} suggest that  Theorem \ref{Thm2} would be applicable 
when there is a ``close spike'' eigenvalue. More detailed discussions are given  in 
Section \ref{Disc}.

We find that in all cases the rLRT has higher empirical power than the cLRT for 
the chosen values of $\lambda=0.4$ and $0.7$.  We also find that  the empirical  power  curve increases 
to 1 less rapidly if $\lambda$ or $\gamma$ is closer to 1. In addition, although we do not report 
the details,  the empirical curves converge fast and do have 
minor changes after $n=80$ for the selected values of $\lambda$
and $\gamma$.

\begin{figure}[htb!]
\centering
\begin{minipage}{.45\textwidth}
  \centering
  \includegraphics[width=1.0\linewidth]{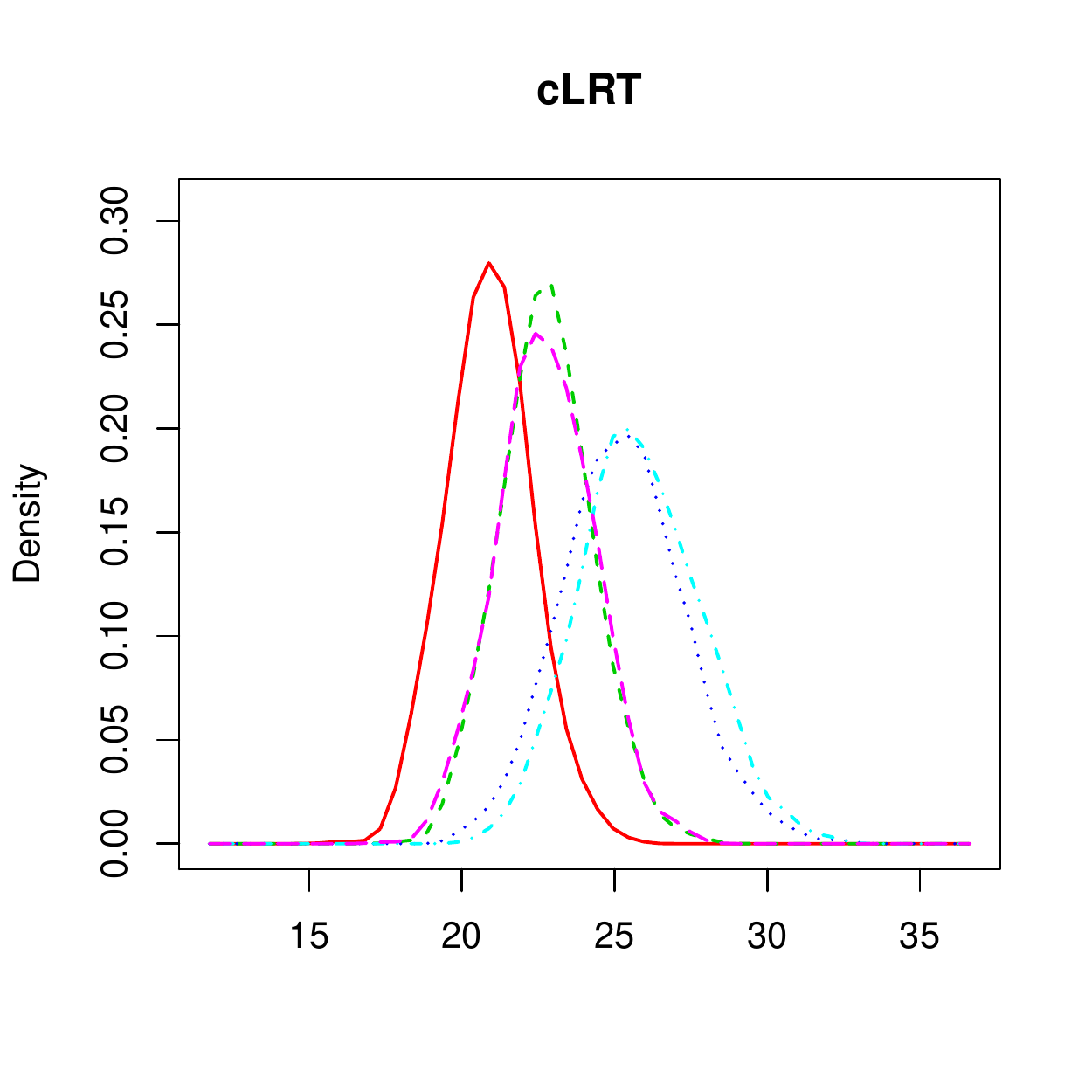}
\end{minipage}%
\begin{minipage}{.45\textwidth}
  \centering
  \includegraphics[width=1.0\linewidth]{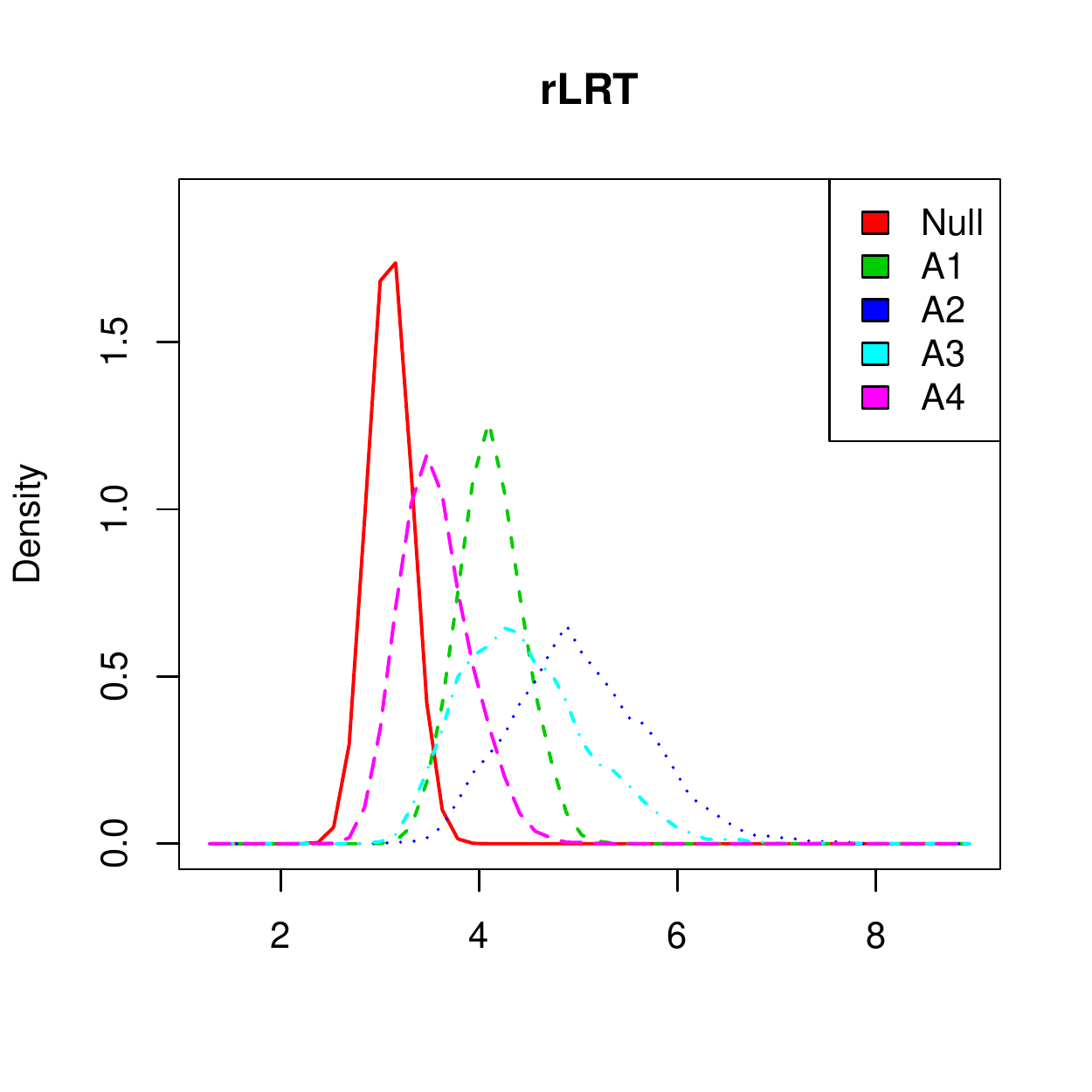}
\end{minipage}%
\caption{{Empirical density functions of the rLRT and the 
cLRT under the null and four alternative hypotheses when $n=40$, $p=32$, and $\lambda=0.5$: (A1) independent but 
heteroscedastic variance case $\Sigma = \diag(2, 2, \cdots , 2, 1, 1, \cdots, 1)$ where the number 
of $2$'s is $\min \{1, \lfloor 0.2p \rfloor \}$, where $\lfloor r \rfloor$ is the round-down of $r$; (A2) independent with a single diverging spiked eigenvalue 
as $\Sigma=\diag(1+0.2p, 1,1,\cdots, 1)$; (A3) compound symmetry $\Sigma = \Sigma_{\rm cs}(0.2)$; and (A4) 
compound symmetry $\Sigma = \Sigma_{\rm cs}(0.1)$ as defined in (\ref{cpsymm}).}}
\label{fig:empipdf}
\end{figure}

To understand the power gain due to the use of  the rLRT better, we plot the empirical 
density of the rLRT and cLRT under {the null and four alternative 
hypotheses (A1)-(A4) (used in Section 4.2) in Figure \ref{fig:empipdf}. }
The figure shows that (i) the variances 
of the rLRT are smaller than the cLRT under both null and alternative hypotheses
and (ii) the distances between the null and the alternative distributions 
are larger in the rLRT than in the cLRT.
Accordingly, the rLRT has larger power than the cLRT and we will see 
this is true regardless of the choice of $n$ and $p$ in the next section.

\subsection{Power comparison with other existing procedures}\label{Num2}

 In this section, we numerically compare the empirical sizes and powers of the rLRT statistic  to other existing tests, namely the corrected LRT (cLRT) by \cite{Bai2009},  the invariant test by \cite{Ledoit2002}, and the non-parametric 
test by \cite{Chen2010}.

In this study, {random samples of size $n$ are generated from the $p-$dimensional multivariate normal distribution ${\rm MVN}_p(0, \Sigma)$. The covariance matrix is set as $\Sigma = {\rm I}_p$ to obtain the empirical sizes.
 The sample size $n$ is chosen as $20, 40, 80$, and $160$, and, for each $n$, $\gamma = \gamma' = p/n$ is
 chosen as $0.2, 0.5$, and $0.8$.} For example, $p = 32, 80$, and $128$ are considered for $n=160$ in the simulation. We take $0.05$ as as the level of significance. 
The shrinkage intensity $\lambda$ of ${\rm rLRT}(\lambda)$ is selected from $0.2, 0.5$, and 
$0.8$ to investigate the effect of the magnitude of the linear shrinkage. This means that we 
compare the empirical sizes of cLRT, rLRT(0.2), rLRT(0.5), and rLRT(0.8) under varying $n$ 
and $\gamma$.

To compare the powers, we consider the following four alternatives: (A1) independent but 
heteroscedastic variance case $\Sigma = \diag(2, 2, \cdots , 2, 1, 1, \cdots, 1)$ where the number 
of $2$'s is $\min \{1, \lfloor 0.2p \rfloor \}$, where $\lfloor r \rfloor$ is the round-down of $r$; (A2) independent with a single diverging spiked eigenvalue 
as $\Sigma=\diag(1+0.2p, 1,1,\cdots, 1)$; (A3) compound symmetry $\Sigma = \Sigma_{\rm cs}(0.2)$; and (A4) 
compound symmetry $\Sigma = \Sigma_{\rm cs}(0.1)$ as defined in (\ref{cpsymm}). Here, the compound symmetry  matrix 
$\Sigma_{\rm cs}(\rho)$ has a single spiked eigenvalue $1+\rho \cdot p$ and $p-1$ non-spiked eigenvalues of $1$. Thus, (A2) and (A3) have the identical spectra.
The sample size $n$ and $\gamma$ are chosen to be the same as those of the null.

The empirical sizes and powers of the listed methods are reported in Table \ref{table:sizepower}. First, the empirical sizes of all the tests approach to the aimed level $0.05$ as $n$ increases. However, the size of \cite{Ledoit2002} shows slower convergence and more upward bias than the size of the other tests do in all cases we considered here. {For this reason, the power of \cite{Ledoit2002} after correcting the size (empirically) are also reported in Table \ref{table:sizepower}, where the cut off value is decided based on $100,000$ simulated test statistics under the null for each simulation setting.}
Second, it can be seen that the emprical powers of the rLRTs are higher than those of the cLRT in all cases we considered. In addition, it is interesting to note that the power improvement is especially higher in the case $\gamma=0.8$ (when $p$ is relatively largfe). 
Third, comparing to the tests of \citet{Chen2010} and (the biased-corrected version of) \citet{Ledoit2002}, the proposed rLRT(0.5) and rLRT(0.2) has higher empirical power in most of the cases. 
Finally, we remark that the computational cost of \citet{Chen2010} is at least $O(pn^4)$ due to the fourth moment calculation so it is not suitable for data with large $n$. In fact, to test the data with $n=500$, \citet{Chen2010} takes tens of hours to finish all computation (using \texttt{C} codes and Intel Core-i7 CPU), whereas the other tests only require seconds.

\begin{table}[hp]
\centering
\bigskip
{\scriptsize
\begin{tabular}[c]{c|c|c|cccc|cc}
\hline
 & $\gamma$ & $n$ & cLRT & 
 $\begin{array}{c}\mbox{rLRT}\\ \mbox{(0.8)}\end{array}$ &
 $\begin{array}{c}\mbox{rLRT}\\ \mbox{(0.5)}\end{array}$ &
 $\begin{array}{c}\mbox{rLRT}\\ \mbox{(0.2)}\end{array}$ &
 Chen & LW \\
\hline
 &     & 20 & 0.085 & 0.078 & 0.078 & 0.083 & 0.087 & 0.089 \\ 
 & 0.2 & 40 & 0.069 & 0.066 & 0.068 & 0.073 & 0.074 & 0.083 \\ 
 &     & 80 & 0.062 & 0.059 & 0.061 & 0.064 & 0.070 & 0.076 \\ 
\cline{2-9} 
 &     & 20 & 0.068 & 0.059 & 0.065 & 0.072 & 0.068 & 0.104 \\ 
Null & 0.5 & 40 & 0.061 & 0.055 & 0.059 & 0.064 & 0.075 & 0.098 \\ 
 &     & 80 & 0.056 & 0.053 & 0.056 & 0.060 & 0.068 & 0.091 \\ 
\cline{2-9} 
 &     & 20 & 0.065 & 0.055 & 0.063 & 0.068 & 0.068 & 0.125 \\ 
 & 0.8 & 40 & 0.058 & 0.052 & 0.056 & 0.059 & 0.058 & 0.120 \\ 
 &     & 80 & 0.055 & 0.051 & 0.054 & 0.056 & 0.064 & 0.113 \\ 
\hline

 &     & 20 & 0.370 & 0.444 & 0.511 & \underline{\bf 0.550} & 0.459 & 0.517 (0.440) \\ 
 & 0.2 & 40 & 0.394 & 0.489 & 0.579 & \underline{\bf 0.632} & 0.540 & 0.596 (0.527) \\ 
A1: &     & 80 & 0.951 & 0.984 & 0.996 & \underline{\bf 0.998} & 0.989 & 0.991 (0.987) \\ 
\cline{2-9} 
Indep. &     & 20 & 0.265 & 0.478 & 0.604 & \underline{\bf 0.653} & 0.478 & 0.574 (0.471) \\ 
with & 0.5 & 40 & 0.569 & 0.880 & 0.961 & \underline{\bf 0.977} & 0.851 & 0.897 (0.849) \\ 
hetero. &     & 80 & 0.967 & 1.000 & 1.000 & 1.000 & 1.000 & 1.000 (0.999) \\ 
\cline{2-9} 
variance &     & 20 & 0.180 & 0.580 & 0.703 & \underline{\bf 0.741} & 0.459 & 0.629 (0.492) \\ 
 & 0.8 & 40 & 0.392 & 0.959 & 0.991 & \underline{\bf 0.995} & 0.854 & 0.924 (0.868) \\ 
 &     & 80 & 0.859 & 1.000 & 1.000 & 1.000 & 0.999 & 1.000 (0.999) \\ 
\hline

 &     & 20 & 0.280 & 0.341 & 0.403 & \underline{\bf 0.440} & 0.351 & 0.408 (0.331) \\ 
A2: & 0.2 & 40 & 0.724 & 0.810 & 0.871 & \underline{\bf 0.899} & 0.853 & 0.882 (0.848) \\ 
Indep. &     & 80 & 0.999 & 1.000 & 1.000 & 1.000 & 1.000 & 1.000 (1.000) \\ 
\cline{2-9} 
with &     & 20 & 0.385 & 0.596 & 0.698 & \underline{\bf 0.739} & 0.664 & 0.729 (0.650) \\ 
single & 0.5 & 40 & 0.890 & 0.979 & 0.993 & \underline{\bf 0.996} & 0.989 & 0.996 (0.994) \\ 
diversing &     & 80 & 1.000 & 1.000 & 1.000 & 1.000 & 1.000 & 1.000 (1.000) \\ 
\cline{2-9} 
spike &     & 20 & 0.349 & 0.765 & 0.838 & \underline{\bf 0.864} & 0.795 & 0.879 (0.816) \\ 
$1+0.2p$ & 0.8 & 40 & 0.867 & 0.997 & 0.999 & \underline{\bf 1.000} & 0.999 & 1.000 (0.999) \\ 
 &     & 80 & 1.000 & 1.000 & 1.000 & 1.000 & 1.000 & 1.000 (1.000) \\ 
\hline

 &     & 20 & 0.280 & 0.343 & 0.404 & \underline{\bf 0.440} & 0.382 & 0.410 (0.333) \\ 
 & 0.2 & 40 & 0.726 & 0.810 & 0.870 & \underline{\bf 0.900} & 0.835 & 0.882 (0.848) \\ 
A3: &     & 80 & 0.999 & 1.000 & 1.000 & 1.000 & 1.000 & 1.000 (1.000) \\ 
\cline{2-9} 
Compound  &     & 20 & 0.387 & 0.595 & 0.695 & \underline{\bf 0.736} & 0.669 & 0.727 (0.649) \\ 
symmetry & 0.5 & 40 & 0.888 & 0.979 & 0.992 & \underline{\bf 0.996} & 0.995 & 0.996 (0.993) \\ 
with &     & 80 & 1.000 & 1.000 & 1.000 & 1.000 & 1.000 & 1.000 (1.000) \\ 
\cline{2-9} 
$\rho=0.2$ &     & 20 & 0.351 & 0.764 & 0.837 & \underline{\bf 0.862} & 0.796 & 0.877 (0.816) \\ 
 & 0.8 & 40 & 0.865 & 0.997 & 0.999 & 1.000 & 0.999 & 1.000 (0.999) \\ 
 &     & 80 & 1.000 & 1.000 & 1.000 & 1.000 & 1.000 & 1.000 (1.000) \\ 
\hline

 &     & 20 & 0.137 & 0.162 & 0.195 & \underline{\bf 0.217} & 0.187 & 0.201 (0.140) \\ 
 & 0.2 & 40 & 0.280 & 0.358 & 0.441 & \underline{\bf 0.491} & 0.400 & 0.454 (0.380) \\ 
A4: &     & 80 & 0.796 & 0.885 & 0.939 & \underline{\bf 0.962} & 0.948 & 0.953 (0.940) \\ 
\cline{2-9} 
Compound &     & 20 & 0.153 & 0.252 & 0.330 & \underline{\bf 0.368} & 0.292 & 0.365 (0.266) \\ 
symmetry & 0.5 & 40 & 0.405 & 0.659 & 0.782 & \underline{\bf 0.830} & 0.772 & 0.828 (0.767) \\ 
with &     & 80 & 0.939 & 0.996 & 0.999 & 1.000 & 1.000 & 1.000 (1.000) \\ 
\cline{2-9} 
$\rho=0.1$ &     & 20 & 0.139 & 0.363 & 0.452 & \underline{\bf 0.490} & 0.403 & 0.526 (0.392) \\ 
 & 0.8 & 40 & 0.376 & 0.838 & 0.911 & \underline{\bf 0.936} & 0.912 & 0.952 (0.920) \\ 
 &     & 80 & 0.915 & 1.000 & 1.000 & 1.000 & 1.000 & 1.000 (1.000) \\ 
\hline 

\end{tabular}
}
\bigskip
\caption{Summary of sizes and powers over 100,000 replications except for \citet{Chen2010} and 1,000 replications for \citet{Chen2010} (due to heavy computation).  The empirically corrected powers of \citet{Ledoit2002} 
are reported in the parentheses. {For each row, the maximum powers are highlighted in bold and the maximum powers among the four LRT-based tests are underlined. The powers for $n=160$ are all equal to $1$ and are removed from the table.}}
\label{table:sizepower}
\end{table}

\section{Discussion}\label{Disc} 

We conclude the paper with a few additional issues of the proposed rLRT not fully discussed in the mainbody of the paper.

First, we consider the case where $n < p$; equivalently, $\gamma \geq 1$. In 
this case, the cLRT is not well-defined because the logarithm term, $\log |{\bf S}_n| = 
\sum_i \log (l_i)$, contains some zero $l_i$'s.
On the other hand, the rLRT is still well-defined as the corresponding logartithm term $\sum_i \log \psi (l_i)$ remains positive even if $\lambda < 1$. Since the CLT of the 
linear spectral statistics holds for $\gamma \in [0, \infty)$ \citep{Bai2004}, it is possible to extend Theorem \ref{Thm1} and \ref{Thm2} in this paper to the case where $\gamma \in [1, \infty)$.

Second, we discuss the case with closely spiked eigenvalues. In the compound symmetry model of (\ref{cpsymm}), the close spiked eigenvalues are those where the spike $1 + \beta$ is smaller than $1+\sqrt{\gamma}$. As shown in Figure \ref{fig:powerCS}, it appears that the power curves over the interval $\beta \in (0, \sqrt{\gamma})$ could be obtained simply by extending the
formula of Corollary 1 to $(0, \sqrt{\gamma})$. We remark that, however, if we follow  
\cite{Wang2013}, the term $\log \psi (1 + \beta)$ in Corollary 1 
should be omitted when $\beta \in (0, \sqrt{\gamma})$, leading to incoherence between the analytical and empirical power curves on $(0, \sqrt{\gamma})$.

Finally, the selection of shrinkage intensity $\lambda$ is still not well understood for hypothesis testing.   As a reviewer points out, when $\lambda$ approaches $0$, the rLRT becomes irrelevant with the alternative covariance matrix and its power is expected to be close to the size. Thus, an appropriate selection of $\lambda$ is 
important for good performance of the rLRT. 
The selection of $\lambda$ for the purpose of improved estimation is well studied in 
the literature, for example, 
\citet{Ledoit2004}, \citet{Schafer2005} and \citet{Warton2008}. However, our additional 
numerical study shows that such a choice of $\lambda$ for the estimation purpose cannot achieve a 
power gain in testing problem. The optimal selection of $\lambda$ for the hypothesis testing needs further research.

\section*{Acknowledgements}

This paper is supported by National Research Foundation of Korea (NRF) grant funded by the government (MSIP) (No. 2011-0030810). The authors are grateful to two reviewers and the AE for a careful reading and providing improvements.

\appendix

\section{Proof of the main theorems}\label{Pf}

\subsection{Proof of Theorem \ref{Thm1}}

Theorem 1 is a direct consequence of Proposition \ref{Prop1}. Here, we calculate the integrals
(\ref{eqn:rmt_mean}) and (\ref{eqn:rmt_var}) for
$$g(z)=g_1(z)=g_2(z)=\psi(z)-\log(\psi(z))-1,$$
where $\psi(z) = \lambda z + (1 - \lambda)$.
\noindent Mean: Using Proposition \ref{Prop1} and the substitution
$$x=1+\gamma-2\sqrt{\gamma}\cos\theta\,,$$
\begin{eqnarray*}
\mu(g_{i}) &=& \frac{g_{i}(a(\gamma))+g_{i}(b(\gamma))}{4}-\frac{1}{2
\pi}  \int_{a(\gamma)}^{b(\gamma)} \frac{g_{i}(x)}{\sqrt{4 \gamma
-(x-1- \gamma)^{2}}} dx\\
&=& \frac{\lambda\gamma-\log\sqrt{(1+\lambda\gamma)^2-4\lambda^2\gamma}}{2}\\
&&-\frac{1}{2\pi}\int^{\pi}_{0} \left( \lambda\gamma-2\lambda\sqrt{\gamma}\cos\theta-\log(1+\lambda\gamma-2\lambda\sqrt{\gamma}\cos\theta) \right)d\theta\\
&=& -\frac{\log\sqrt{(1+\lambda\gamma)^2-4\lambda^2\gamma}}{2}+\frac{1}{2\pi}\int^{\pi}_{0} \left( \log(1+\lambda\gamma-2\lambda\sqrt{\gamma}\cos\theta) \right)d\theta\\
&=& -\frac{\log\sqrt{(1+\lambda\gamma)^2-4\lambda^2\gamma}}{2}+\frac{1}{4\pi}\int^{2\pi}_{0} \left( \log(1+\lambda\gamma-2\lambda\sqrt{\gamma}\cos\theta) \right)d\theta\,.
\end{eqnarray*}

\noindent Variance: We write $m_1:=\underline{m}(z_1)$ and
$m_2=\underline{m}(z_2)$ for notational simplicity. We have
\begin{equation}\label{eqn:var}
v(g)=-\frac{1}{2\pi^{2}} \oint g(z_{2}(m_2))\oint
\frac{g(z_{1}(m_1))}{(m_1-m_2)^{2}}
dm_1 dm_2,
\end{equation}
To evaluate this integral with Cauchy's formula, we need to identify the points of singularity in $g(z(m_1))\,.$ It can be seen that there is singularity when $\psi(z(m_1))=0\,.$ Rewrite $\psi(z(m))$ as
$$\psi(z(m_1))=\lambda\left( -\frac{1}{m_1}+\frac{\gamma}{m_1+1} \right)+1-\lambda\\
=\frac{(1-\lambda)(m_1-M)(m_1-N)}{m_1(m_1+1)}$$
where
$$M,N=\frac{-(1-2\lambda+\lambda\gamma)\pm\sqrt{(1-2\lambda+\lambda\gamma)^2+4\lambda(1-\lambda)}}{2(1-\lambda)}\,.$$
Then, $M,N$ are points of singularity. Next, choose contours $\mathcal{C}_1$ and $\mathcal{C}_2$ enclosing $-1$ and $M\,,$ but not $0$ and $N\,,$ such
that on the contours, the logarithm in $g(z)$ is single-valued. In addition,
$\mathcal{C}_1$ and $\mathcal{C}_2$ are chosen so that they do not overlap. Applying integration by parts and Cauchy's formula, we have
\begin{eqnarray}
&& \oint \frac{g(z(m_1))}{(m_{1}-m_{2})^{2}}dm_{1}\nonumber\\
&=& \oint \left\{ \frac{\lambda}{m_1^2}-\frac{\lambda\gamma}{(m_1+1)^2}+\frac{1}{m_1}+\frac{1}{m_1+1}-\frac{1}{m_1-M}-\frac{1}{m_1-N}
\right\} \frac{1}{m_{1}-m_{2}}dm_{1} \nonumber\\
&=& 2\pi i \left\{  \frac{\lambda\gamma}{(m_2+1)^2}-\frac{1}{m_2+1}+\frac{1}{m_2-M} \right\}
\label{eqn:second-1}
\end{eqnarray}
Then,
\begin{eqnarray*}
v(g)&=&-\frac{1}{2\pi^{2}} \oint g(z_{2}(m_2))\oint
\frac{g(z_{1}(m_1))}{(m_1-m_2)^{2}}
dm_1 dm_2\\
&=&-\frac{i}{\pi} \oint g(z_{2}(m_2))
\left\{  \frac{\lambda\gamma}{(m_2+1)^2}-\frac{1}{m_2+1}+\frac{1}{m_2-M} \right\}
 dm_2\,.
\end{eqnarray*}
Here,
\begin{eqnarray}
&& \oint \frac{g(z(m_2))}{(m_{2}+1)^{2}}dm_{2}\nonumber\\
&=& \oint \left\{ \frac{\lambda}{m_2^2}-\frac{\lambda\gamma}{(m_2+1)^2}+\frac{1}{m_2}+\frac{1}{m_2+1}-\frac{1}{m_2-M}-\frac{1}{m_2-N}
\right\} \frac{1}{m_2+1}dm_{2} \nonumber\\
&=&  \frac{2\pi i}{1+N}
\label{eqn:second-2}
\end{eqnarray}
Applying integration by parts, Cauchy's formula, and Lemma \ref{contour}, we have
\begin{eqnarray}
&& \oint \frac{g(z(m_1))}{(m_{1}-m_{2})^{2}}dm_{1}\nonumber\\
&=& \oint \left\{ \frac{\lambda}{m_1^2}-\frac{\lambda\gamma}{(m_1+1)^2}+\frac{1}{m_1}+\frac{1}{m_1+1}-\frac{1}{m_1-M}-\frac{1}{m_1-N}
\right\} \frac{1}{m_{1}-m_{2}}dm_{1} \nonumber\\
&=& 2\pi i \left\{  \frac{\lambda\gamma}{(m_2+1)^2}-\frac{1}{m_2+1}+\frac{1}{m_2-M} \right\}
\label{eqn:second-1}
\end{eqnarray}
Then,
\begin{eqnarray*}
v(g)&=&-\frac{1}{2\pi^{2}} \oint g(z_{2}(m_2))\oint
\frac{g(z_{1}(m_1))}{(m_1-m_2)^{2}}
dm_1 dm_2\\
&=&-\frac{i}{\pi} \oint g(z_{2}(m_2))
\left\{  \frac{\lambda\gamma}{(m_2+1)^2}-\frac{1}{m_2+1}+\frac{1}{m_2-M} \right\}
 dm_2\,.
\end{eqnarray*}
Using integration by parts,
\begin{eqnarray}
&& \oint \frac{g(z(m_2))}{(m_{2}+1)^{2}}dm_{2}\nonumber\\
&=& \oint \left\{ \frac{\lambda}{m_2^2}-\frac{\lambda\gamma}{(m_2+1)^2}+\frac{1}{m_2}+\frac{1}{m_2+1}-\frac{1}{m_2-M}-\frac{1}{m_2-N}
\right\} \frac{1}{m_2+1}dm_{2} \nonumber\\
&=& 2\pi i\left( \lambda-1+\frac{1}{1+N} \right)\,.
\label{eqn:second-2}
\end{eqnarray}
Applying Lemma \ref{contour} and Cauchy's formula, we obtain
\begin{eqnarray}
\oint \frac{g(z(m_2))}{m_{2}+1}dm_{2}
= - 2\pi i \left\{ \log (1-\lambda)(1+N)\right\}
\label{eqn:second-3}
\end{eqnarray}
and
\begin{eqnarray}
\oint \frac{g(z(m_2))}{m_{2}-M}dm_{2}
= 2\pi i \left\{ -\frac{\lambda}{M}-\lambda-\log\frac{(1-\lambda)(M-N)}{M} \right\}
\label{eqn:second-4}
\end{eqnarray}
Combining the results (\ref{eqn:second-1}) - (\ref{eqn:second-4}), we have the desired result of variance. \qed

\subsection{Proof of Theorem \ref{Thm2}}

We compute
\begin{equation} \nonumber 
 \int g(x) dF^{\tilde{\gamma}', H_p}(x)  = \int \big\{ \lambda x - \lambda - \log(\lambda x + 1 - \lambda) \big\}  dF^{\gamma', H_p}(x),
\end{equation} 
where $H_p$ is the SD of spiked population model, which is written as
\begin{equation} \nonumber
H_p(t) = \frac{p - K}{p}\delta_1(t) + \frac{1}{p}\sum_{i=1}^{K}{n_i \delta_{a_i}(t)}. 
\end{equation} 
Following the lines of Section 3 in \cite{Wang2013},
\begin{equation} \nonumber
  \int \big( \lambda x - \lambda \big) dF^{\tilde{\gamma}', H_p}(x) = 
\frac{\lambda}{p} \sum_{i=1}^{K} n_i a_i - \frac{\lambda K}{p} + O(\frac{1}{n^2}).  
\end{equation} 
The difficult part lies on the evaluation of integration of the logarithm-related term. Using the labeling in Proposition \ref{Prop2}, we rewrite it as
\begin{equation} \nonumber 
 \int \log(\lambda x + 1 - \lambda) dF^{\tilde{\gamma}', H_p}(x) = (\ref{eqn:spiketerm1}) + (\ref{eqn:spiketerm2}) + (\ref{eqn:spiketerm3})
\end{equation} 
and calculate (\ref{eqn:spiketerm1}), (\ref{eqn:spiketerm2}) and (\ref{eqn:spiketerm3}) separately.
In the remainder of the proof, we write $m$ and $\gamma$ instead of $\underline{m}$ and $\tilde{\gamma}'$, respectively, for notational convenience. The terms (\ref{eqn:spiketerm1}) and (\ref{eqn:spiketerm2}) involve  contour integrals. Recall that the contour $\mathcal{C}$ on (\ref{eqn:spiketerm1}) and (\ref{eqn:spiketerm2}) encloses the closed interval $[ \frac{-1}{1-\sqrt{\gamma}}, \frac{-1}{1+\sqrt{\gamma}}]$ on the real axis of the complex plane and has poles of $\{m=-1\}$, $\{m=M\}$, where $M=M(\lambda, \gamma)$ is defined in (\ref{MN}).
It is easy to show that  $\frac{-1}{1-\sqrt{\gamma}} < M < \frac{-1}{1+\sqrt{\gamma}}$ and $N > 0$ provided $\gamma \in (0,1)$ and $n$ is large, where $N=N(\lambda, \gamma)$ is from (\ref{MN}).

Following the lines of Section 3.3 of \cite{Wang2013}, recall that $\lambda (- \frac{1}{m} + \frac{\gamma}{1+m} ) + (1 - \lambda ) = \frac{(1-\lambda)(m-M)(m-N)}{m(m+1)}$. We have 
\begin{eqnarray*}
&& (\ref{eqn:spiketerm1}) \\
&=& \frac{-1}{2 \pi i p} \oint_{\mathcal{C}} \log\Big( \lambda (- \frac{1}{m} + \frac{\gamma}{1+m} ) + (1 - \lambda )\Big) \cdot \Big( \frac{K}{\gamma m} - \sum_{i=1}^{K}{\frac{n_i a_i^2 m}{(1 + a_i m)^2}} \Big) dm
\\
&=& \frac{-1}{2 \pi i p \gamma} \oint_{\mathcal{C}} \frac{ \log\Big( \lambda (- \frac{1}{m} + \frac{\gamma}{1+m} ) + (1 - \lambda ) \Big) }{m} \cdot \Big( K - \sum_{i=1}^{K}{\frac{n_i a_i^2 m^2 \gamma}{(1 + a_i m)^2}} \Big) dm \\
\\
&=& \frac{-1}{2 \pi i p \gamma} \oint_{\mathcal{C}} \frac{ \log\big( \frac{(1-\lambda)(m-N)}{m} \big) + \log\big( \frac{m-M}{m+1} \big) }{m} \cdot \Big( K - \sum_{i=1}^{K}{\frac{n_i a_i^2 m^2 \gamma}{(1 + a_i m)^2}} \Big) dm
\\
&=& \frac{-K}{2 \pi i p \gamma} \oint_{\mathcal{C}} \frac{\log\big( \frac{m-M}{m+1} \big)}{m} dm +  \frac{1}{2 \pi i p \gamma} \oint_{\mathcal{C}} \log\big( \frac{m-M}{m+1} \big) \sum_{i=1}^{K}{\frac{n_i a_i^2 m \gamma}{(1 + a_i m)^2}} dm
\\
&\triangleq& A_1 + A_2.
\end{eqnarray*}
Here, 
\begin{eqnarray*}
A_1 &=&  \frac{-K}{2 \pi i p \gamma} \oint_{\mathcal{C}} \log\big( \frac{m-M}{m+1} \big) d \log m 
\\ 
&=&  \frac{K}{2 \pi i p \gamma} \oint_{\mathcal{C}} \log m \cdot d \log \big( \frac{m-M}{m+1} \big) 
\\ 
&=&  \frac{K}{2 \pi i p \gamma} \cdot (M+1) \cdot \oint_{\mathcal{C}} \frac{ \log m } {(m+1)(m-M)}dm
\\
&=&  \frac{K}{p \gamma} \log(-M),
\end{eqnarray*}
and 
\begin{eqnarray*}
A_2 &=&  \frac{1}{2 \pi i p } \oint_{\mathcal{C}} \log\big( \frac{m-M}{m+1} \big) \sum_{i=1}^{K}{\frac{n_i a_i^2 m }{(1 + a_i m)^2}} dm
\\
&=& \frac{1}{2 \pi i p } \sum_{i=1}^{K} \oint_{\mathcal{C}} \log\big( \frac{m-M}{m+1} \big) \cdot n_i a_i \Big( \frac{1}{1+a_i m} - \frac{1}{(1 + a_i m)^2} \Big) dm
\\
&\triangleq& A_3 - A_4, 
\end{eqnarray*}
where
\begin{eqnarray*}
A_3 &=& \frac{1}{2 \pi i p } \sum_{i=1}^{K} \oint_{\mathcal{C}}  \log\big( \frac{m-M}{m+1} \big) \frac{n_i a_i}{1 + a_i m} dm
\\
&=& \frac{1}{2 \pi i p } \sum_{i=1}^{K} \oint_{\mathcal{C}} n_i \log\big( \frac{m-M}{m+1} \big) d \log(1 + a_i m)
\\
&=& \frac{-1}{2 \pi i p } \sum_{i=1}^{K} \oint_{\mathcal{C}} n_i \log(1 + a_i m) \cdot
d\log\big( \frac{m-M}{m+1} \big)
\\
&=& \frac{-1}{2 \pi i p } \cdot (M+1) \sum_{i=1}^{K} \oint_{\mathcal{C}} \frac { n_i \log(1 + a_i m) } { (m+1)(m-M)}dm
\\
&=& \frac{1}{p} \sum_{i=1}^{K}{n_i \log(1-a_i)} - \frac{1}{p} \sum_{i=1}^{K}{n_i \log(1 + a_iM)},
\end{eqnarray*}
and
\begin{eqnarray*}
A_4 &=& \frac{1}{2 \pi i p } \sum_{i=1}^{K} \oint_{\mathcal{C}} \log\big( \frac{m-M}{m+1} \big) \frac{n_i a_i}{(1 + a_i m)^2} dm
\\
&=& \frac{1}{2 \pi i p } \sum_{i=1}^{K} \oint_{\mathcal{C}}  \frac{n_i}{1 + a_i m} \cdot
d \log \big( \frac{m-M}{m+1} \big)
\\
&=& \frac{M+1}{2 \pi i p } \sum_{i=1}^{K} \oint_{\mathcal{C}} \frac{n_i}{(1+ a_i m)(m - M)(m + 1)} dm
\\
&=& \frac{1}{p} \sum_{i=1}^{K} n_i \big( \frac{1}{1 + a_i M} - \frac{1}{1 - a_i} \big).
\end{eqnarray*}
Combiing the results of $A_1 + A_2 = A_1 + A_3 - A_4$, we have
\begin{equation} \nonumber 
(\ref{eqn:spiketerm1}) = \frac{K}{p \gamma} \log(-M) + \frac{1}{p} \sum_{i=1}^{K}{n_i \log \Big( \frac{1-a_i}{1 + a_iM} \Big)} 
 - \frac{1}{p} \sum_{i=1}^{K} n_i \Big( \frac{1}{1 + a_i M} - \frac{1}{1 - a_i} \Big).
\end{equation}

Next, consider (\ref{eqn:spiketerm2}). Taking $f=\log \circ \psi$, we have
\begin{equation} \nonumber 
(\log \circ \psi )'\big( -\frac{1}{m} + \frac{\gamma}{1+m} \big)
 = \frac{\lambda}{\lambda x + (1 - \lambda)} \Bigg|_{x= -\frac{1}{m} + \frac{\gamma}{1+m}}
 = \frac{\lambda m (m+1) }{(1-\lambda)(m-M)(m-N)}. 
\end{equation} 
Then
\begin{eqnarray*}
&& (\ref{eqn:spiketerm2}) \\
&=& -\frac{1}{2 \pi i p} \oint_{\mathcal{C}} (\log \circ \psi )'\big( -\frac{1}{m} + \frac{\gamma}{1+m} \big) \sum_{i=1}^{K} \big( \frac{n_i a_i}{1 + a_i m} - \frac{n_i}{1+m} \big) \big( \frac{1}{m} - \frac{\gamma m}{(1+m)^2} \big)dm
\\
&=& -\frac{1}{2 \pi i p} \cdot \frac{\lambda}{1-\lambda} \sum_{i=1}^{K}  n_i \oint_{\mathcal{C}}  \frac{ m (m+1) }{(m-M)(m-N)} \big( \frac{ a_i}{1 + a_i m} - \frac{1}{1+m} \big) \big( \frac{1}{m} - \frac{\gamma m}{(1+m)^2} \big)dm
\\
&=&  -\frac{1}{2 \pi i p} \cdot \frac{\lambda}{1-\lambda} \sum_{i=1}^{K} n_i (B_1 - B_2 - B_3 + B_4),
\end{eqnarray*}
where
\begin{eqnarray*}
B_1 &=& \oint_{\mathcal{C}} \frac{a_i (m+1)}{(m-M)(m-N)(1 + a_i m)} dm  \\ 
&=& 2 \pi i \cdot \bigg( \frac{a_i (M+1)}{(1 + a_i M)(M-N)} \bigg),
\\
B_2 &=& \oint_{\mathcal{C}} \frac{a_i \gamma m^2}{(m-M)(m-N)(1 + a_i m)(m+1)} dm 
\\ 
&=& 2 \pi i \cdot \bigg( \frac{a_i \gamma M^2}{(M-N)(1+a_iM)(M+1)} + \frac{a_i \gamma}{(M+1)(N+1)(1-a_i)}  
 \bigg),
\\
B_3 &=& \oint_{\mathcal{C}} \frac{1}{(m-M)(m-N)} dm = 2 \pi i \cdot \frac{1}{M-N},
\end{eqnarray*}
and
\begin{eqnarray*}
B_4 &=& \oint_{\mathcal{C}} \frac{ \gamma m^2}{(m-M)(m-N)(m+1)^2} dm
\\
&=& 2 \pi i \cdot \bigg( \frac{\gamma M^2}{(M-N)(M+1)^2} - \frac{\gamma (2MN + M + N)}{(M+1)^2(N+1)^2} \bigg).
\end{eqnarray*}
Collecting the four terms, we have :
\begin{eqnarray*}
&& (\ref{eqn:spiketerm2}) \\
&=& -\frac{\lambda}{p(1 - \lambda)} \cdot \sum_{i=1}^{K} n_i \Bigg[ \frac{1}{M-N} \Bigg\{ \frac{a_i (M+1)}{1+ a_i M} -\frac{a_i \gamma M^2}{(1 + a_i M)(M+1)} - 1 + \frac{\gamma M^2}{(M+1)^2} \Bigg\} 
\\
&&\qquad\qquad\qquad\qquad\qquad - \frac{1}{(M+1)(N+1)} \Bigg\{ \frac{a_i \gamma}{1 - a_i} + \frac{ \gamma (2MN+M+N) }{(M+1)(N+1)} \Bigg\} \Bigg].
\end{eqnarray*}

To obtain (\ref{eqn:spiketerm3}), note that the integration term $ \int \log(\psi(x)) dF^{\gamma, \delta_1}(x)$ is equal to $- \int \{ \psi(x) + \log(\psi(x)) - 1 \} dF^{\gamma, \delta_1}(x)$ since M-P law satisfies $ \int x dF^{\gamma, \delta_1}(x) =  \int 1 dF^{\gamma, \delta_1}(x)$. This gives
\begin{equation} \nonumber 
(\ref{eqn:spiketerm3}) = -\left( 1 - \frac{K}{p} \right)  \int g(x) dF^{\gamma, \delta_1}(x) + \frac{1}{p} \sum_{i=1}^{K} n_i \log \psi \{ \varphi(a_i) \} + O(\frac{1}{n^2})
\end{equation}
where $\varphi(a_i) = a_i + \frac{\gamma a_i}{a_i -1}$.

Finally, combining the four results, we finally obtain the centering term :
\begin{eqnarray*}
&& \int \big\{ \psi(x) - \log(\psi(x)) - 1 \big\} dF^{\gamma, H_p}(x) \\
&=& \int \big( \lambda x - \lambda \big) dF^{\gamma, \delta_1}(x) - (\ref{eqn:spiketerm1}) - (\ref{eqn:spiketerm2}) - (\ref{eqn:spiketerm3}).
\\
&=& \left( 1 - \frac{K}{p} \right)  \int g(x) dF^{\gamma, \delta_1}(x) + \frac{1}{p} C (\lambda, \gamma  + O(\frac{1}{n^2}),
\end{eqnarray*}
where
\begin{eqnarray*}
C_n &=& \lambda \cdot \frac{1}{K} \sum_{i=1}^{K} n_i a_i - \lambda  -  \frac{1}{K}\sum_{i=1}^{K} n_i \log \psi \{ \varphi(a_i) \} \\
&& \hspace*{-1.2cm} - \Bigg[ \frac{1}{ \gamma} \log(-M) + \frac{1}{K}\sum_{i=1}^{K}n_i \log \Big( \frac{1-a_i}{1 + a_i M} \Big) 
- \frac{1}{K} \sum_{i=1}^{K} n_i \Big( \frac{1}{1 + a_i M} - \frac{1}{1 - a_i} \Big)  \Bigg] \\
&& \hspace*{-1.2cm} + \frac{\lambda}{(1 - \lambda)K} \sum_{i=1}^{K} n_i \Bigg[ \frac{1}{M-N} \Bigg\{ \frac{a_i (M+1)}{1+ a_i M} -\frac{a_i \gamma M^2}{(1 + a_i M)(M+1)} - 1 + \frac{\gamma M^2}{(M+1)^2} \Bigg\} 
\\
&& \hspace*{-1.2cm}~~~~~~~~~~~~~\qquad\qquad - \frac{1}{(M+1)(N+1)} \Bigg\{ \frac{a_i \gamma}{1 - a_i} + \frac{\gamma (2MN + M + N)}{(M+1)(N+1)} \Bigg\} \Bigg]. \qed
\end{eqnarray*}

\subsection{Technical lemma}

\begin{lemma} Let $z_A$ and $z_B$ be any two different fixed complex numbers. Then, (a) for any contour $\mathcal{C}$ enclosing
$z_A$ and $z_B$ such that $$\log\frac{z-z_A}{z-z_B}$$
is single-valued on the contour, we have
$$\oint \frac{1}{z-z_A}\log\frac{z-z_A}{z-z_B}dz=0\,,$$
(b) for any contour $\mathcal{C}$ enclosing
$z_A$ and $z_B\,,$ we have
$$\oint \frac{dz}{(z-z_A)(z-z_B)}=0$$
\label{contour} \end{lemma}

\begin{figure}[tbh]
\label{msft} {\small \vspace{-5mm} } 
\par
\begin{center}
$
\begin{array}{cc}
\includegraphics[width=110mm,height=60mm]{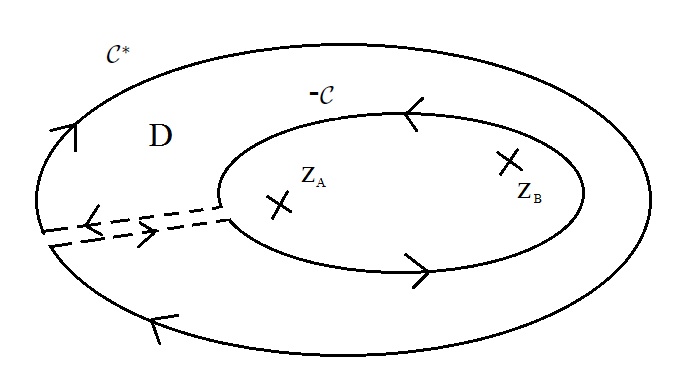} & 
\end{array}
$%
\end{center}
\par
{\small \vspace{-5mm} }
\caption{Illustration of the path of the contour integral}
\end{figure}

\begin{proof}
 (a) Let $\mathcal{C}^*$ be a contour enclosing $\mathcal{C}$ such that both $\mathcal{C}$ and $\mathcal{C}^*$ are clockwise (anticlockwise), and on $\mathcal{C}^*\,,$ we have $|z-z_A|<|z_B-z_A|$ (see Figure 3). Then, $D\,,$ the singly connected region between $\mathcal{C}$ and $\mathcal{C}^*$ as indicated in Figure 3 contains no singularity. Therefore, the integral on $\mathcal{C}$ and $\mathcal{C}^*$ are the same. Next, consider the power series expansion
\begin{eqnarray*}
\log\frac{z-z_A}{z-z_B}&=&-\log\left( 1+\frac{z_A-z_B}{z-z_A} \right)\\
&=&-\left\{ \left(\frac{z_A-z_B}{z-z_A}\right)-\frac{1}{2}\left(\frac{z_A-z_B}{z-z_A}\right)^2+\frac{1}{3}\left(\frac{z_A-z_B}{z-z_A}\right)^3-\ldots \right\}\,.
\end{eqnarray*}
Such power series converges on $\mathcal{C}^*\,.$ The desired result is a consequence of
$$\oint \frac{dz}{(z-z_A)^i}=0$$
for all $i=2,3,\ldots\,.$

\noindent (b) Applying Cauchy's formula, we have
\begin{equation} \nonumber 
\oint \frac{dz}{(z-z_A)(z-z_B)}
=\frac{1}{z_A-z_B}\oint \left\{\frac{1}{z-z_A}-\frac{1}{z-z_B}\right\}dz
=\frac{1-1}{z_A-z_B}=0.
\end{equation} 

\end{proof}

\end{document}